\documentclass[aip,reprint]{revtex4-1}

\usepackage{amsfonts}
\usepackage{amsmath}
\usepackage{amssymb}
\usepackage{booktabs}
\usepackage{caption}
\usepackage{color}
\usepackage{comment}
\usepackage{graphicx}
\usepackage{hyperref}
\usepackage{subfig}
\usepackage{xspace}

\newcommand{\pygbe}{\texttt{PyGBe}\xspace}
\newcommand{\gb}{{\small G\,B1\,D4$^\prime$}\xspace}
\newcommand{\ig}{{\small IgG}}
\newcommand{\pdb}{{\small PDB}\xspace}
\newcommand{\gmres}{\textsc{gmres}\xspace}
\newcommand{\bem}{\textsc{bem}\xspace}
\newcommand{\ses}{\textsc{ses}\xspace}
\newcommand{\sam}{\textsc{sam}}
\newcommand{\gpu}{\textsc{gpu}}
\newcommand{\cpu}{\textsc{cpu}}

\newcommand{\nvidia}{\textsc{nvidia}\xspace}
\newcommand{\msms}{\texttt{\textsc{msms}}\xspace}
\newcommand{\amber}{\texttt{\textsc{amber}}\xspace}
\newcommand{\ccby}{\textsc{cc-by}\xspace}

\graphicspath{{figs/}} %  PATH to figure files-- change to ./ for submission

\begin{document}

%\preprint{}

\title{Probing protein orientation near charged nanosurfaces for simulation-assisted biosensor design}


\author{Christopher D. Cooper}
\email[]{cdcooper@bu.edu,christopher.cooper@usm.cl}
%\thanks{}
\affiliation{Mechanical Engineering, Boston University, Boston, MA}
\affiliation{Mechanical Engineering, Universidad T\'ecnica Federico Santa Mar\'ia, Valpara\'iso, Chile}

\author{Natalia C. Clementi}
\email[]{ncclementi@gwu.edu}
\affiliation{Mechanical \& Aerospace Engineering, The George Washington University, Washington, DC.}

\author{Lorena A. Barba}
\email[]{labarba@gwu.edu}
\homepage[]{http://lorenabarba.com/}
\affiliation{Mechanical \& Aerospace Engineering, The George Washington University, Washington, DC.}


\date{\today}

\begin{abstract}

Protein-surface interactions are ubiquitous in biological processes and bioengineering, yet are not fully understood. 
In biosensors, a key factor determining the sensitivity and thus the performance of the device is the orientation of the ligand molecules on the bioactive device surface. Adsorption studies thus seek to determine how orientation can be influenced by surface preparation, varying surface charge and ambient salt concentration.
In this work, protein orientation near charged nanosurfaces is obtained under electrostatic effects using the Poisson-Boltzmann equation, in an implicit-solvent model.
Sampling the free energy for protein \gb at a range of tilt and rotation angles with respect to the charged surface, we calculated the probability of the protein orientations and observed a dipolar behavior. This result is consistent with published experimental studies and combined Monte Carlo and molecular dynamics simulations using this small protein, validating our method. 
More relevant to biosensor technology, antibodies such as immunoglobulin G are still a formidable challenge to molecular simulation, due to their large size. 
With the Poisson-Boltzmann model, we obtained the probability distribution of orientations for the iso-type {\small IgG2a} at varying surface charge and salt concentration.
This iso-type was not found to have a preferred orientation in previous studies, unlike the iso-type {\small IgG1} whose larger dipole moment was assumed to make it easier to control.  
Our results show that the preferred orientation of {\small IgG2a} can be favorable for biosensing with positive charge on the surface of 0.05C/m$^{2}$ or higher and 37mM salt concentration.
The results also show that local interactions dominate over dipole moment for this protein.
Improving immunoassay sensitivity may thus be assisted by numerical studies using our method (and open-source code), guiding changes to fabrication protocols or protein engineering of ligand molecules to obtain more favorable orientations. 

\end{abstract}


\maketitle %\maketitle must follow title, authors, abstract and \pacs

% Body of paper.

\section{Introduction}\label{sec:intro}
%!TEX root = CooperBarba-orientation.tex

Protein adsorption plays a key role in many biotechnological applications, particularly biomaterials and tissue engineering, biomedical implants and biosensors.
Yet, despite their importance, the specific mechanisms governing protein-surface interactions are not fully understood.\cite{Gray2004,RabeVerdesSeegel2011}

In the field of biosensors, protein adsorption needs to be engineered to obtain a successful device. 
Biosensors detect specific molecules using a nanoscale sensing element, such as a metallic nanoparticle or nanowire covered with a bioactive coating. 
The prevailing method to modify a sensor surface is via a self-assembled monolayer (\sam) of a small charged group, with ligand molecules layered on top to achieve the desired function. 
Antibodies are a common choice for the ligand molecules, although the newest devices use single-domain or single-chain fragment molecules.\cite{ByunETal2013,TrillingETal2014} 
Sensing occurs when a target biomolecule binds to the ligand molecule,  changing some physical parameter on the sensor, such as current in nanowires or plasmon resonance frequency in metallic nanoparticles. 

One of the factors crucially affecting biosensor performance is the orientation of ligand molecules.\cite{TajimaTakaiIshihara2011,TrillingBeekwilderZuilhof2013} 
These have specific binding sites, which need to be accessible to the target molecule for the biosensor to function well.
Probing protein orientation is thus one key goal of adsorption studies.
The aim of this study is to ascertain how orientation can be influenced by fabrication conditions regarding surface preparation, such as surface charge and ambient salt content. We consider in particular the antibody immunoglobulin G near a solid surface at different charge concentrations and ionic strengths. Using a smaller molecule (protein \gb), we could first confirm agreement of our results with published works reporting experiments\cite{BaioWeidnerBaughGambleStaytonCastner2012} and  simulations with a combined Monte Carlo and molecular dynamics method.\cite{LiuLiaoZhou2013} These previous works, among others, also concluded that electrostatic interactions are the dominant effect in the orientation of adsorbed proteins. In the case of immunoglobulin G (\ig), the protein is relevant for biosensor applications, but its large size would make all-atom molecular simulations quite cumbersome and expensive. For this reason, other researchers have studied adsorption of \ig\ using a coarse-grained model that considers each residue as a sphere (united-residue model),\cite{ZhouChenJiang2003} finding that electrostatics dominates the orientation for higher surface charges and that a positive charge can result in the desired ``tail-on'' placement for the \ig 1 iso-type, at low enough salt concentration. 
Here, we investigate the preferred orientations for the \ig 2a variant, which other researchers found hard to control.
In addition to obtaining the preferred orientation at different conditions of charge and ionic strength, we also take a detailed look at the probability distribution in the parameter space.

Our model for protein-surface interactions uses the Poisson-Boltzmann equations in their integral formulation, representing the protein geometry as a dielectric interface in an implicit solvent. We recently verified the model against an analytical solution valid for spherical geometries and studied its numerical convergence in detail.\cite{CooperBarba2015a}
Previous studies on protein-surface interaction using the Poisson-Boltzmann equation showed that such a model is adequate as long as conformational changes in the protein are slight,\cite{YaoLenhoff2004,YaoLenhoff2005} and also that van der Waals effects can be neglected for realistic molecular geometries.\cite{RothNealLenhoff1996}
Conformational changes of the biomolecule can be ignored in this case because binding sites need to remain nearly unmodified during the biosensor fabrication process.\cite{TajimaTakaiIshihara2011} 
A continuum framework has been used in the past to study protein orientation,\cite{JufferArgosDevlieg1996} but it included ions explicitly. Other studies have used a coarse-grained model of the molecule, represented as  a set of spheres,\cite{ShengTsaoZhouJiang2002,ZhouTsaoShengJiang2004} assigned effective charges at the residue level,\cite{FreedCramer2011,ZhouChenJiang2003} or made approximations to account for pH effects.\cite{BiesheuvelvanderVeenNord2005,HartvigdeWeertOstergaartJorgensenJensen2011}

The sensor element (functionalized with the \sam) is represented in our model as a charged surface that interacts electrostatically with the biomolecule. A parameter sweep of the protein's rotation and tilt angles with respect to the solid surface provides energy landscapes, where the probability of finding the system in a given micro-state depends on the total free energy.
The continuum approach can thus provide insights to the conditions (surface charge and salt concentration) conducive to a favorable orientation of large proteins, too large for all-atom molecular simulation with today's computing power. It can also represent solid surfaces of any geometry, and we expect that it may in future assist in the design of better ligand-molecule immobilization techniques for high-sensitivity biosensors.

\section{Implicit-solvent model for proteins near charged surfaces} \label{sec:implicit_solvent}
%!TEX root = CooperBarba-orientation.tex

The implicit-solvent model describes a molecular system as a set of continuum dielectric regions, and computes the mean-field potential using electrostatics. 
For the case where a protein is dissolved in a solvent, we require two of such regions: inside and outside the protein, interfaced by the solvent-excluded surface (\ses). 
The \ses determines the closest a water molecule can get to the protein, and we generate it by rolling a spherical probe of the size of a water molecule around the protein. 
The dielectric constant inside the protein is low ($\epsilon= 2\text{ to }4$) and there are point charges  placed at the atomic locations. The solvent region has the dielectric constant of water $\epsilon \approx 80$, and we need to account for the presence of salt. 
This model results in a system of partial differential equations where the Poisson equation describes the electrostatic potential inside the protein, and the linearized Poisson-Boltzmann equation applies outside the protein. On the \ses, appropriate interface conditions ensure the continuity of the potential and electric displacement.

\begin{figure}[h]
   \includegraphics[width=0.45\textwidth]{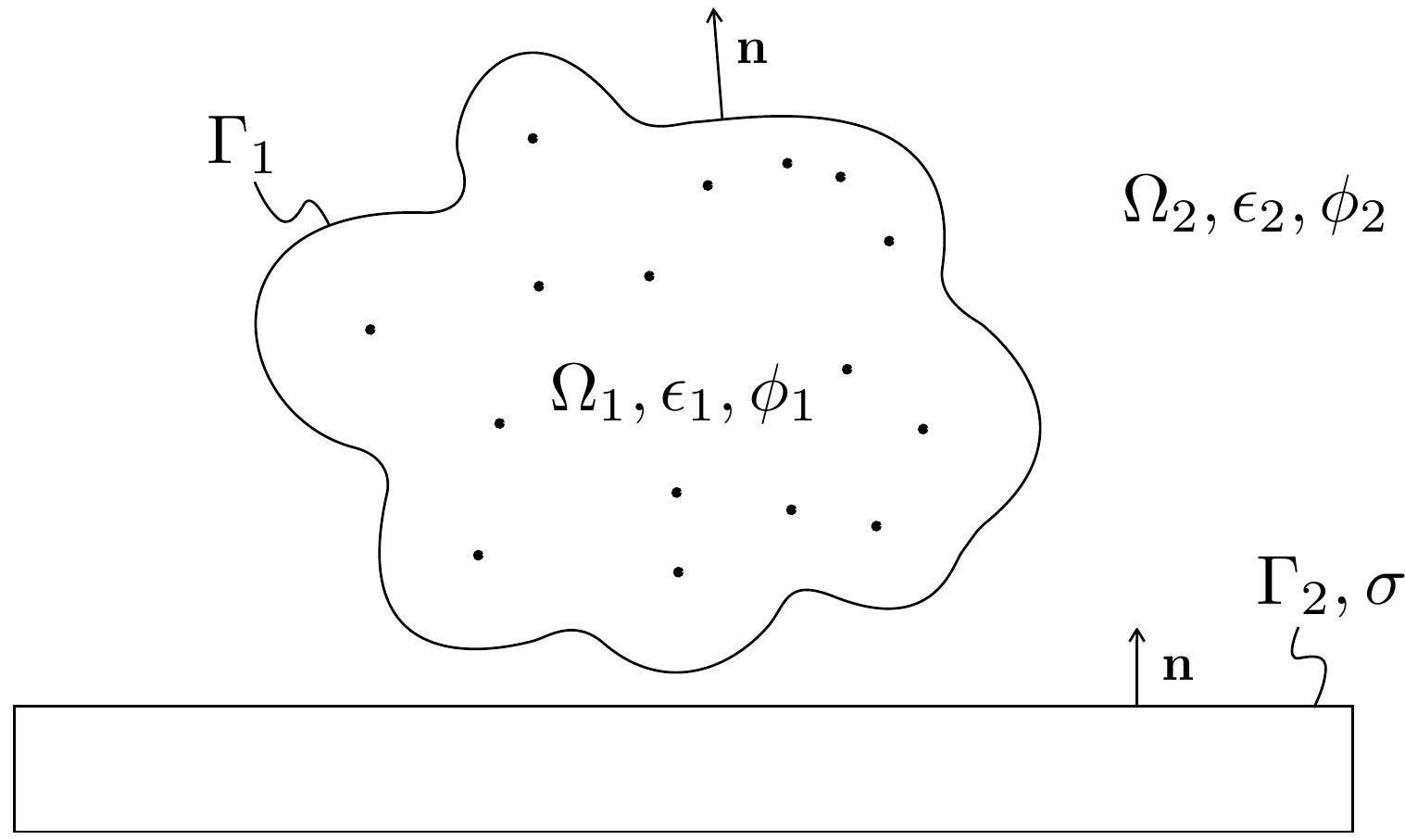} 
   \caption{Sketch of a molecule interacting with a surface: $\Omega_1$ is the protein, $\Omega_2$ the solvent region, $\Gamma_1$ is the  \ses and $\Gamma_2$ a surface with imposed charge.}
   \label{fig:molecule_surface}
\end{figure}

In this work, we use an extension of the implicit-solvent model to consider the effect of charged surfaces. Such is the case of the setup sketched by Figure \ref{fig:molecule_surface}, which is described mathematically by the following equations:

\begin{align} \label{eq:pde}
\nabla^2 \phi_1(\mathbf{r}) &= - \sum_k \frac{q_k}{\epsilon_1} \delta(\mathbf{r},\mathbf{r}_k) \ \text{ in solute $(\Omega_1)$,}  \nonumber \\ 
\nabla^2\phi_2 (\mathbf{r}) &= \kappa^2 \phi_2(\mathbf{r}) \quad \qquad \ \ \ \text{ in solvent $(\Omega_2)$,}  \nonumber \\ 
\phi_1 &=\phi_2 \nonumber \\ 
\epsilon_1 \frac{\partial \phi_1}{\partial \mathbf{n}} &= \epsilon_2 \frac{\partial \phi_2}{\partial \mathbf{n}}  \ \qquad \qquad \text{ on interface $\Gamma_1$, and} \nonumber \\
-\epsilon_2 \frac{\partial \phi_2}{\partial \mathbf{n}} &= \sigma_0 \qquad \qquad \qquad \text{ on surface $\Gamma_2$} 
\end{align}

\noindent where $\phi_i$ is the electrostatic potential in region $\Omega_i$, which has a permittivity $\epsilon_i$, and $\sigma_0$ is a prescribed charge on the surface. The surface $\Gamma_2$ could correspond to a device such as a biosensor.

\paragraph*{Boundary integral formulation---} \label{sec:bie}
%!TEX root = CooperBarba-orientation.tex

We apply Green's second identity to the system of partial-differential equations in \eqref{eq:pde}, and evaluate the resulting equations on $\Gamma_1$ and $\Gamma_2$ to obtain the following system of integral equations:
\begin{widetext}
\begin{align} \label{eq:integral_eq}
\frac{\phi_{1,\Gamma_1}}{2}+ K_{L}^{\Gamma_1}(\phi_{1,\Gamma_1}) -  V_{L}^{\Gamma_1} \left(\frac{\partial}{\partial \mathbf{n}}\phi_{1,\Gamma_1} \right)  =  
\frac{1}{\epsilon_1} \sum_{k=0}^{N_q} \frac{q_k}{4\pi|\mathbf{r}_{\Gamma_1} - \mathbf{r}_k|} &  \quad \text{on $\Gamma_1$,} \nonumber \\ 
\frac{\phi_{1,\Gamma_1}}{2} - K_{Y}^{\Gamma_1}(\phi_{1,\Gamma_1}) +  \frac{\epsilon_1}{\epsilon_2} V_{Y}^{\Gamma_1} \left( \frac{\partial}{\partial \mathbf{n}} \phi_{1,\Gamma_1} \right) -  
K_{Y}^{\Gamma_1}(\phi_{2,\Gamma_2})  + V_{Y}^{\Gamma_1} \left( -\frac{\sigma_0}{\epsilon_2} \right)  = 0& \quad \text{on $\Gamma_1$,} \nonumber \\ 
- K_{Y}^{\Gamma_2}(\phi_{1,\Gamma_1}) + \frac{\epsilon_1}{\epsilon_2} V_{Y}^{\Gamma_2}  \left( \frac{\partial}{\partial \mathbf{n}} \phi_{1,\Gamma_1} \right) + \frac{\phi_{2,\Gamma_2}}{2} - 
K_{Y}^{\Gamma_2}(\phi_{2,\Gamma_2}) +  V_{Y}^{\Gamma_2} \left( -\frac{\sigma_0}{\epsilon_2} \right)  = 0& \quad \text{on $\Gamma_2$.}
\end{align}
\end{widetext}

\noindent The function $\phi_{i,\Gamma_j} = \phi_i(\mathbf{r}_{\Gamma_j})$ is the electrostatic potential at a point that approaches the surface $\Gamma_j$ from the region $\Omega_i$, and
$K$ and $V$ are known as the single- and double-layer potentials, correspondingly:
\begin{align} \label{eq:layers}
K_{L/Y}^{\Gamma_k}(\phi_{i,\Gamma_j}) &= \oint_{\Gamma_j} \frac{\partial}{\partial \mathbf{n}} \left[ G_{L/Y}(\mathbf{r}_{\Gamma_k},\mathbf{r}_{\Gamma_j}) \right]\phi_{i,\Gamma_j} \, \mathrm{d} \Gamma, \nonumber \\
V_{L/Y}^{\Gamma_k} \left( \frac{\partial}{\partial \mathbf{n}} \phi_{i,\Gamma_j} \right) &= \oint_{\Gamma_j} \frac{\partial}{\partial \mathbf{n}} \phi_{i,\Gamma_j} G_{L/Y}(\mathbf{r}_{\Gamma_k},\mathbf{r}_{\Gamma_j})  \, \mathrm{d} \Gamma.
\end{align}

\noindent In Equation \eqref{eq:layers}, $G_L$ and $G_Y$ are the free-space Green's functions of the Poisson and linearized Poisson-Boltzmann equations, respectively. 
The single-layer potential of a distribution $\psi$ on a surface $\Gamma$ evaluated at $\mathbf{r}$, $V^\mathbf{r}(\psi_\Gamma)$, can be interpreted as the potential on $\mathbf{r}$ due to a charge distribution $\psi$ on $\Gamma$. 
Similarly, $K^\mathbf{r}(\psi_\Gamma)$ can be seen as the potential induced by a double layer of charges ($\psi$) with opposite sign at $\Gamma$.

Rearranging terms, we write Equation \eqref{eq:integral_eq} in matrix form, as follows:
 \begin{align} \label{eq:matrix_dphi}
 \left[
    \begin{matrix} % or pmatrix or bmatrix or Bmatrix or ...
       \frac{1}{2} + K_{L}^{\Gamma_1} & -V_{L}^{\Gamma_1} & 0 \vspace{0.2cm}\\
       \frac{1}{2} - K_{Y}^{\Gamma_1} &  \frac{\epsilon_1}{\epsilon_2} V_{Y}^{\Gamma_1} & -K_{Y}^{\Gamma_1} \vspace{0.2cm} \\
       - K_{Y}^{\Gamma_2} & \frac{\epsilon_1}{\epsilon_2} V_{Y}^{\Gamma_2} & \left(\frac{1}{2} - K_{Y}^{\Gamma_2}\right) \\
    \end{matrix}
    \right] \left[ 
    \begin{matrix} % or pmatrix or bmatrix or Bmatrix or ...
       \phi_{1,\Gamma_1} \vspace{0.2cm} \\
       \frac{\partial}{\partial \mathbf{n}} \phi_{1,\Gamma_1} \vspace{0.2cm}\\
       \phi_{2,\Gamma_2}\\
    \end{matrix} 
     \right] =   \nonumber \\
    \left[
    \begin{matrix} % or pmatrix or bmatrix or Bmatrix or ...
       \sum_{k=0}^{N_q} \frac{q_k}{4\pi|\mathbf{r}_{\Gamma_1} - \mathbf{r}_k|} \vspace{0.2cm} \\
        V_{Y}^{\Gamma_1} \left( \frac{\sigma_0}{\epsilon_2} \right) \vspace{0.2cm} \\
        V_{Y}^{\Gamma_2} \left( \frac{\sigma_0}{\epsilon_2} \right)
    \end{matrix}
    \right].
 \end{align}

The boundary-integral formulation is not limited to represent the protein with a single surface, but can account for solvent-filled cavities inside the protein region and Stern layers.\cite{CooperBardhanBarba2013} In those cases, more than one surface is required to appropriately represent the protein. 
Our implementation follows the guidelines from Altman and co-workers\cite{AltmanBardhanWhiteTidor09} to deal with multiple surfaces.

The boundary-integral formulation of the implicit-solvent model is a popular alternative to compute solvation energies of proteins,\cite{YoonLenhoff1990, Juffer1991a, LuETal2006, BajajETal2011, AltmanBardhanWhiteTidor09, GengKrasny2013, CooperBardhanBarba2013} but the effect of charged surfaces has rarely been considered. The only work that we know of that does include these effects is limited to plane, infinite surfaces.\cite{YoonLenhoff1992}

%=============
\section{Methods}\label{sec:methods}
%!TEX root = CooperBarba-orientation.tex

\subsection{Discretization and implementation details}

We solve the system in \eqref{eq:integral_eq} numerically using a boundary element method (\bem). To represent the \ses, we use flat triangular panels where the potential $(\phi)$ and its normal derivative $(\partial \phi /\partial \mathbf{n})$ are constant, and then collocate the discretized equation on the center of each panel. This transforms the integral operators in the matrix equation \eqref{eq:matrix_dphi} into block matrices of size $N_p \times N_p$, where $N_p$ is the number of panels. Each entry of the block matrix is an integral over one panel ($\Gamma_j$), evaluated on the center of panel $\Gamma_i$:
\begin{align} \label{eq:layers_element}
K_{L,ij} &= \int_{\Gamma_j} \frac{\partial}{\partial \mathbf{n}} \left[ G_L(\mathbf{r}_{\Gamma_i},\mathbf{r}_{\Gamma_j}) \right]\mathrm{d} \Gamma_j, \nonumber \\
V_{L,ij} &= \int_{\Gamma_j} G_L(\mathbf{r}_{\Gamma_i},\mathbf{r}_{\Gamma_j})  \mathrm{d} \Gamma_j.
\end{align}

We classify the integrals in Equation \eqref{eq:layers_element} in three groups, depending on the distance $d$ between the panel and the collocation point. 
When the collocation point is inside the panel being integrated, we get a singular integral that we solve with a semi-analytical approach\cite{ZhuHuangSongWhite2001} placing Gauss nodes on the sides of the triangle. 
We call near-singular integrals those where $d<2L$ ($L = \sqrt{2\cdot \text{Area}}$). For near-singular integrals, we use a high-order Gauss quadrature rule with 19 or more nodes. 
Finally, when the panel and collocation points are further than $2L$ from each other, we only need 1, 3, 4 or 7 Gauss nodes per element to get good accuracy.

To solve the resulting linear system, we use a general minimal residual method (\gmres). The most time consuming part of the \gmres solver is a matrix-vector multiplication---in principle, an $O(N^2)$ operation---done within every iteration of the solver. But by using a treecode algorithm, we perform this operation in $O(N\log N)$ time.\cite{BarnesHut1986} More details on our implementation of the \bem can be found in our earlier work,\cite{CooperBarba-share154331} and a companion paper.\cite{CooperBarba2015a}
 %discretization and treecode

\subsection{Energy calculation} \label{sec:energy}
%!TEX root = CooperBarba-orientation.tex

We can decompose the total free energy into Coulombic, surface, and solvation energy:

\begin{equation}
F_\text{Total} = F_\text{Coulomb} + F_\text{surf} + F_\text{solv}.
\end{equation}

\medskip

\paragraph*{Coulombic energy---}

The Coulombic energy arises simply from the Coulomb interactions of all point charges. We compute it by

\begin{equation} \label{eq:coul_energy}
F_\text{Coulomb} = \frac{1}{2} \sum_j^{N_q}\sum^{N_q}_{\substack{i\\ i\neq j}} q_iq_j\frac{1}{4\pi |\mathbf{r}_i-\mathbf{r}_j|}
\end{equation}

\paragraph*{Solvation free energy---}

The solvation energy is the energy contribution of the protein's surroundings: solvent polarization, charged surfaces, and other proteins. We compute it as

\begin{align} \label{eq:solv_energy}
F_{\text{solv}} &= \frac{1}{2} \int_{\Omega} \rho \,(\phi_{\text{total}} - \phi_{\text{Coulomb}}) \\
&= \sum_{k=0}^{N_q} q_k (\phi_{\text{total}} - \phi_{\text{Coulomb}})(\mathbf{r}_k),
\end{align}

\noindent where $\rho$ is the charge distribution, consisting of point charges (which transforms the integral into a sum), and $\phi_\text{reac} = \phi_{\text{total}} - \phi_{\text{Coulomb}}$ is
\begin{equation} \label{eq:phi_reac_bem}
\phi_{\text{reac},\mathbf{r}_k} = -K_{L}^{\mathbf{r}_k}(\phi_{1,\Gamma_1}) + V_{L}^{\mathbf{r}_k} \left(\frac{\partial}{\partial \mathbf{n}}\phi_{1,\Gamma_1} \right) 
\end{equation}

\paragraph*{Surface free energy---}

We use the description of free energy of a surface with prescribed charge (like $\Gamma_2$ in Figure \ref{fig:molecule_surface}) from Chan and co-workers.\cite{ChanMitchell1983,CarnieChan1993} They describe the free energy on a surface as

\begin{equation} \label{eq:energy_surf}
F_\text{surf} = \frac{1}{2} \int_{\Gamma} G_c \sigma_0^2 d\Gamma, 
\end{equation} 

\noindent where $\phi = G_c \sigma_0$.

\subsection{Orientation sampling of a protein near a charged surface}  \label{sec:prot_orientation}
%!TEX root = CooperBarba-orientation.tex

\begin{figure}%[h] 
   \centering
   \includegraphics[width=0.5\textwidth]{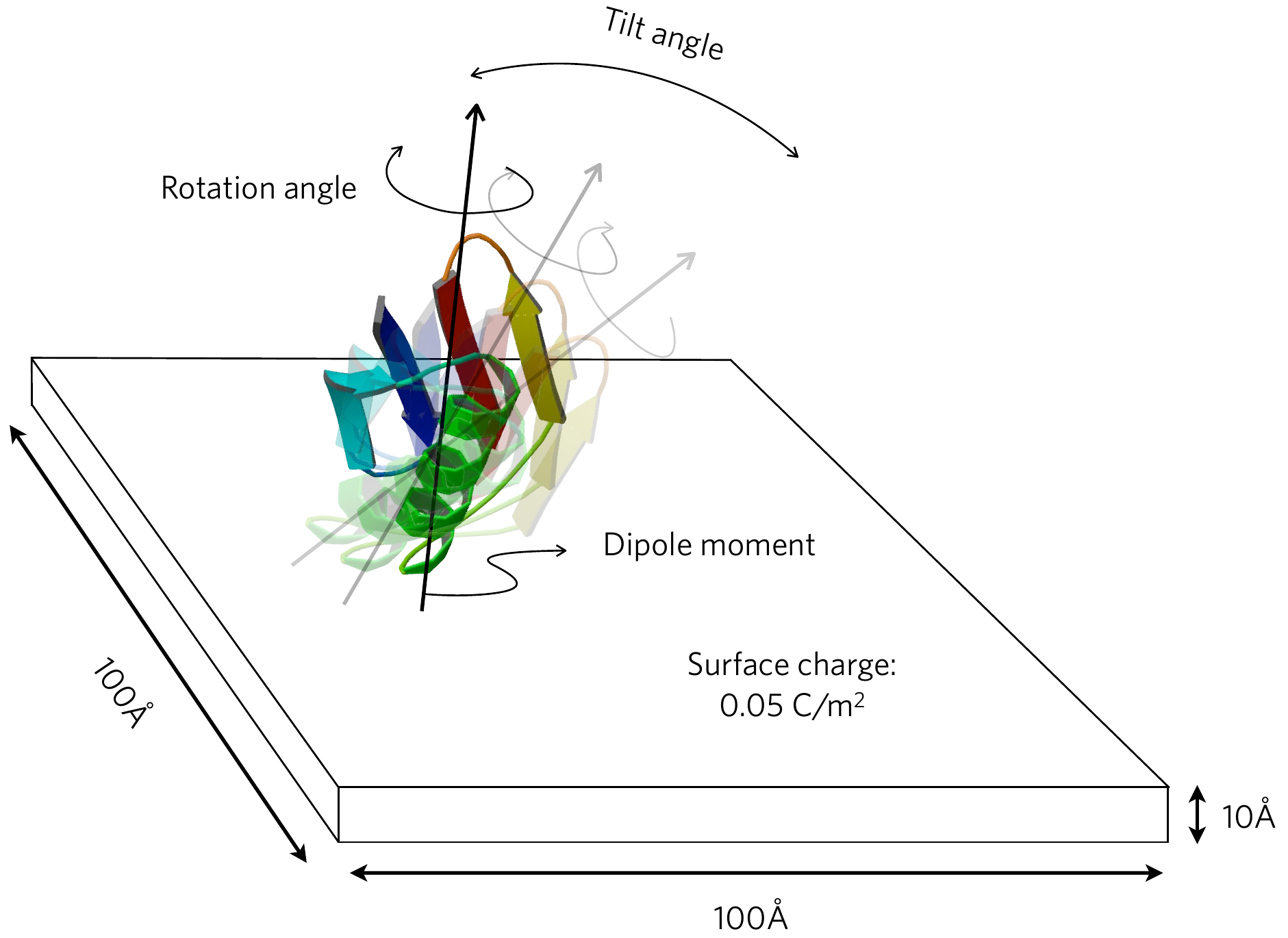}
   \caption{Setup of the problem for our orientation-sampling studies.}
   \label{fig:1pgb_orientation}
\end{figure}

We are aiming to investigate the orientation of proteins near self-assembled monolayers (\sam), specifically for biosensing applications. In the framework of the implicit-solvent model, we can represent the \sam\ as a surface charge density, and use Equation \eqref{eq:matrix_dphi} to compute the electrostatic potential. 
According to the Boltzmann distribution, the probability of finding the system in micro-state $\lambda$ depends on the total free energy, $F_\text{total}$, as follows:
\begin{equation} \label{eq:prob}
P(\lambda) = \frac{\int_{\lambda} \exp \left(-\frac{F_\text{total}}{k_B T} \right) \text{d} \lambda}{\int_{\Lambda} \exp \left(-\frac{F_\text{total}}{k_B T} \right) \text{d} \Lambda},
\end{equation} 

\noindent where $\Lambda$ is the ensemble of all micro-states, $k_B$ is the Boltzmann constant and $T$ the temperature. To obtain a probability distribution, we assume that electrostatic effects are dominant and use Equation \eqref{eq:prob}, sampling $F_\text{total}$ for different orientations. We define the orientation using the angle between the dipole moment and surface normal vectors as a reference (tilt angle), varying from 0$^\circ$ to 180$^\circ$. For each tilt angle, we rotate the protein about the dipole moment vector by 360$^\circ$ to examine all possible orientations. This process is sketched in Figure \ref{fig:1pgb_orientation}.

In this case, micro-states are defined by the tilt ($\alpha_{\text{tilt}}$) and rotational ($\alpha_{\text{rot}}$) angles, and we rewrite the integral in the numerator of Equation \eqref{eq:prob} as:
\begin{equation} \label{eq:prob_angle}
\int_{\lambda} \exp \left(-\frac{F_\text{total}}{k_B T} \right) \text{d} \lambda = \int \int \exp \left(-\frac{F_\text{total}}{k_B T} \right) \text{d} \alpha_{\text{rot}} \text{d} \alpha_{\text{tilt}},
\end{equation}

\noindent where micro-state $\lambda$ is a range of angles $\alpha_{\text{rot}}$ and $\alpha_{\text{tilt}}$. 
In biosensors, the ligand is adsorbed on the surface (usually covalently), hence we are interested on the orientation of the molecule very close to the surface, and don't consider configurations away from it.

\subsection{Structure preparation}

\begin{figure}%[h] 
   \centering
   \includegraphics[width=0.25\textwidth]{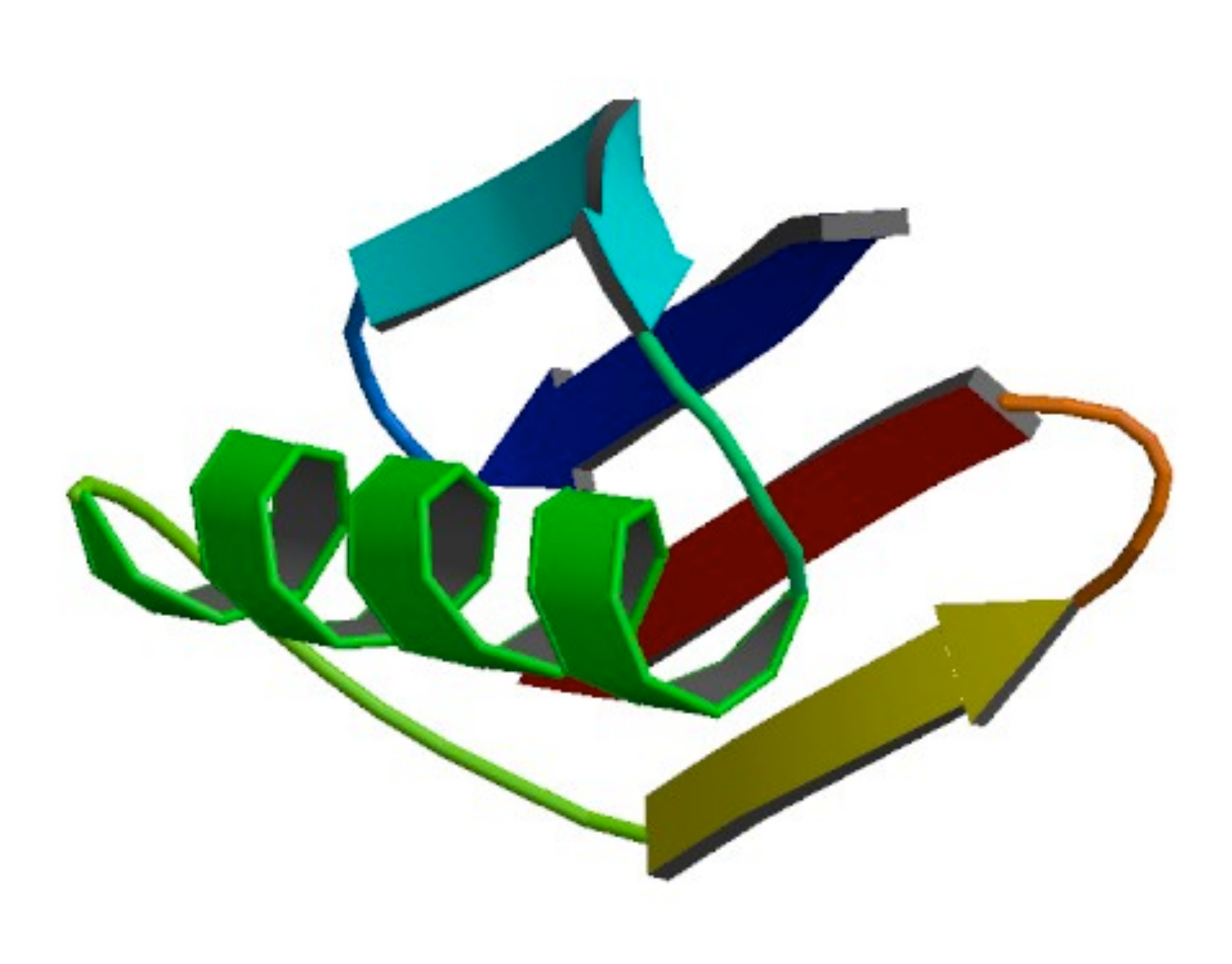}
   \caption{Structure of protein \gb (\pdb code {\small 1PGB}).}
   \label{fig:1pgb}
\end{figure}

\begin{figure}%[h] 
   \centering
   \includegraphics[width=0.35\textwidth]{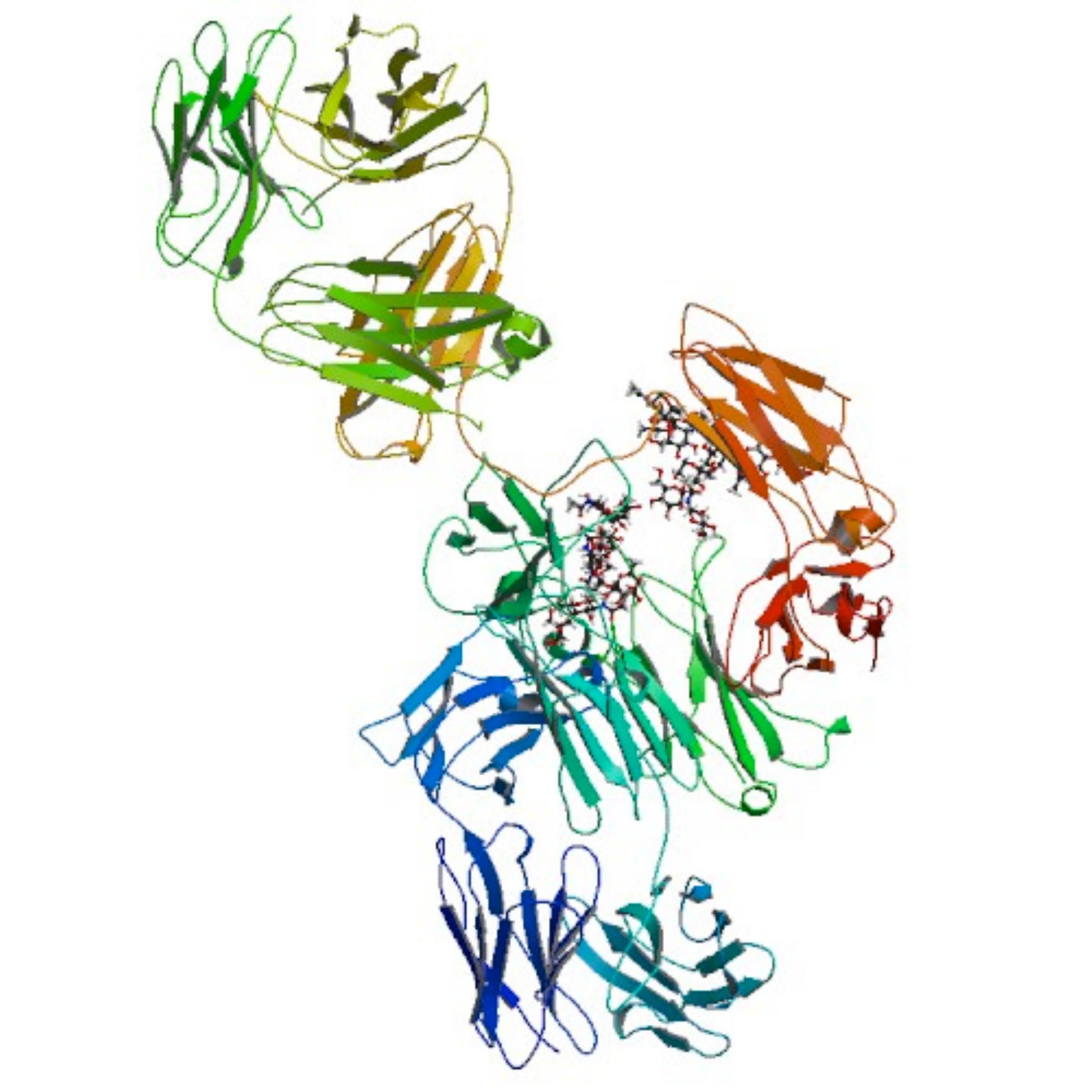}
   \caption{Structure of immunoglobulin G (\pdb code {\small 1IGT}).}
   \label{fig:1igt}
\end{figure}

To assess the adequacy of the implicit-solvent model for investigating protein-surface interactions, we studied the orientation of protein \gb mutant near a charged surface, since there are results available in the literature that we could compare to: both experimental observations \cite{BaioWeidnerBaughGambleStaytonCastner2012} and simulations using a combined Monte Carlo and molecular dynamics approach.\cite{LiuLiaoZhou2013} Figure \ref{fig:1pgb} shows the structure of protein \gb (\pdb code {\small 1PGB}), to which we applied mutations {\small E19Q}, {\small D22N}, {\small D46N} and {\small D47N} to obtain the {\small D4$^\prime$} mutant, using the \textsl{SwissPdb Viewer} software.\cite{GuexPeitsch1997}
We then studied the orientation of antibody immunoglobulin G iso-type \ig 2a (\pdb code {\small 1IGT}), a widely used protein in biosensors, whose structure is shown in Figure \ref{fig:1igt}. This is a more interesting case from the point of view of our application, yet it is more difficult to study with molecular simulation due to its size.
In both cases, the vector orientation of the dipole moment (used as reference for the tilt and orientation angles) was obtained using the location of the point charges at the locations of the atoms.

%=============

\section{Results} \label{sec:results}
%!TEX root = CooperBarba-orientation.tex

The results detailed in this section were obtained using an extension of the open-source code \pygbe,\footnote{\href{https://github.com/barbagroup/pygbe}{https://github.com/barbagroup/pygbe}} accounting for the presence of surfaces with imposed charge or potential.\cite{CooperBarba2015a}
We ran the calculations for protein \gb on a workstation with Intel Xeon X5650 \cpu s  and one \nvidia Tesla C2075 \gpu\ card (late 2011). 
The second case considers the antibody immunoglobulin G, which is a much larger molecule than protein \gb. For these runs, we used either: (1) Boston University's \textsc{bungee} cluster, which has 16 nodes with 8 Intel Xeon \cpu\ cores each, and a total of 3 \nvidia Tesla K20 (Kepler, late 2012) and 26 \nvidia Tesla M2070/2075 \gpu s; 
or (2) the George Washington University's \textsl{Colonial One} cluster, with 32 \gpu\ nodes featuring dual Intel Xeon E2620 2.0GHz 6-core processors with dual \nvidia K20 and 128 GB of memory.
All runs were serial: single-\cpu\ and single-\gpu. 
We obtained the van der Waals radii and charge distribution using \texttt{pdb2pqr}\cite{Dolinsky04} with an \amber forcefield, and generated the meshes using the free \msms software.\cite{SannerOlsonSpehner1995}
In these tests, we did not consider a Stern layer for either the protein or the charged surface, nor the presence of solvent-filled cavities inside the protein.

\subsection{First case: protein G\,B1\,D4$^{\prime}$} \label{sec:PGB}

We investigated the preferred orientation of protein \gb placed 2\AA\ away from a 100\AA$\times$100\AA$\times$10\AA\ block with surface charge density $\pm$0.05C/m$^2$, centered with respect to a 100\AA$\times$100\AA\ face.
In biosensors, protein \gb can be used as an intermediate protein, coupled to the functionalized surface directly by covalent bonding.
The protein will thus be at a small distance from the surface. In this case, 2\AA\ is in the order of magnitude of the size of a water molecule, or of a C--N bond.
The charge density of $\pm$0.05C/m$^2$ matches that used in other works.\cite{LiuLiaoZhou2013}
In these cases, we considered a solvent with no salt, i.e., $\kappa=0$ (to compare with other published results), and with relative permittivity 80. The region inside the protein had a relative permittivity of 4.

As seen in Figure \ref{fig:1pgb_orientation}, $\alpha_\text{tilt}$ is the angle between the protein's dipole moment and the normal vector to the surface, and $\alpha_\text{rot}$ rotates about the dipole moment. 
When the dipole-moment vector and the normal are aligned ($\alpha_\text{tilt}=0$), we define a vector $\mathbf{V}_\text{ref}$ as the shortest distance between the axis normal to the surface that goes through the center of mass, and the atom that is furthest away from it. 
We use $\mathbf{V}_\text{ref}$ as a reference to define the rotation angle $\alpha_\text{rot}$: the angle between $\mathbf{V}_\text{ref}$ and the vector normal to a 100\AA$\times$10\AA\ face.

We sampled the total free energy every $\Delta \alpha_{\text{tilt}} = 2^\circ$ of tilt angle and $\Delta \alpha_{\text{rot}} =10^\circ$ of rotation angle, resulting in $3,240$ independent runs.  The surface mesh had 4 triangles per square Angstrom on the protein geometry and 2 triangles per square Angstrom on the charged surface. 

Numerical parameters are presented in Table \ref{table:params3}. In a companion publication,\cite{CooperBarba2015a} we present a grid-convergence study using both an analytical solution and a case with protein \gb.\cite{CooperBarba2015-share1348803} We computed an approximate exact value of $-222.43$[kcal/mol] for solvation energy and $317.98$[kcal/mol] for surface energy using Richardson extrapolation with very fine parameters. With results that are less than 2\% away from the approximate exact values, we are comfortable with the parameters in Table \ref{table:params3} and mesh densities of $4$ elements per square Angstrom on the protein and $2$ elements per square Angstrom on the surface.\cite{CooperBarba2015a}

\begin{table}[h]
  %\centering
   %\fontfamily{ppl}\selectfont
   \caption{\label{table:params3}Numerical parameters used for numerically probing the orientation of protein \gb. } 
    \begin{tabular}{c c c c c c c}
	\hline%\toprule
	\multicolumn{3}{l} {\# Gauss points:} & \multicolumn{3}{l}{Treecode:} & \gmres:\\
	\footnotesize{in-element} & \footnotesize{close-by} & \footnotesize{far-away} & $N_{\text{crit}}$ & $P$ &  $\theta$  & tol.\\
	\hline%\midrule
	9 per side & 19 & 1  &  300 & 4 & 0.5  & $10^{-5}$\\	
	\hline%\bottomrule
    \end{tabular}
\end{table}

\begin{figure*}
   \centering
   \subfloat[]{\includegraphics[width=0.43\textwidth]{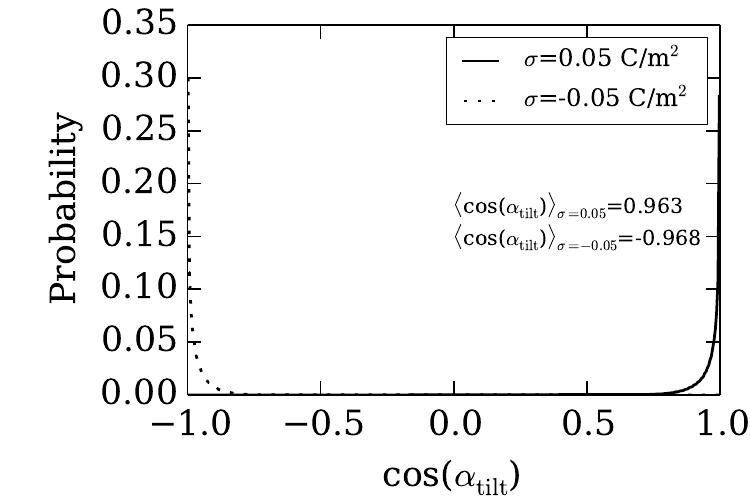} \label{fig:1PGB_cos}}
   \subfloat[]{\includegraphics[width=0.43\textwidth]{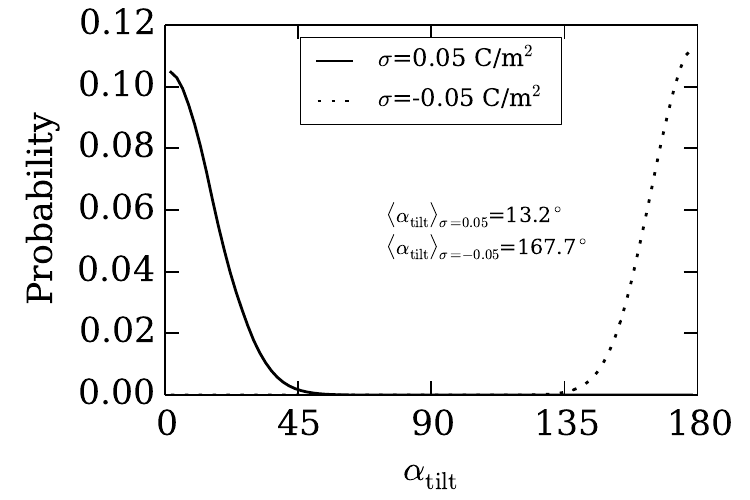} \label{fig:1PGB_alpha}}\\
   \subfloat[]{\includegraphics[width=0.40\textwidth]{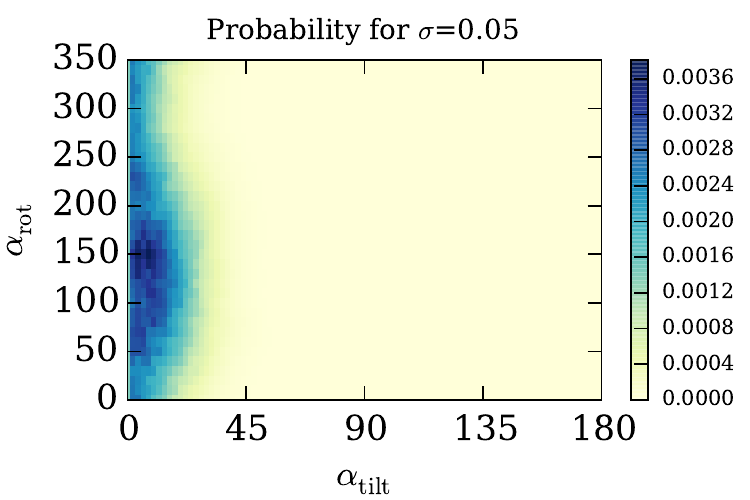} \label{fig:1PGB_2D_sig005}}
   \subfloat[]{\includegraphics[width=0.40\textwidth]{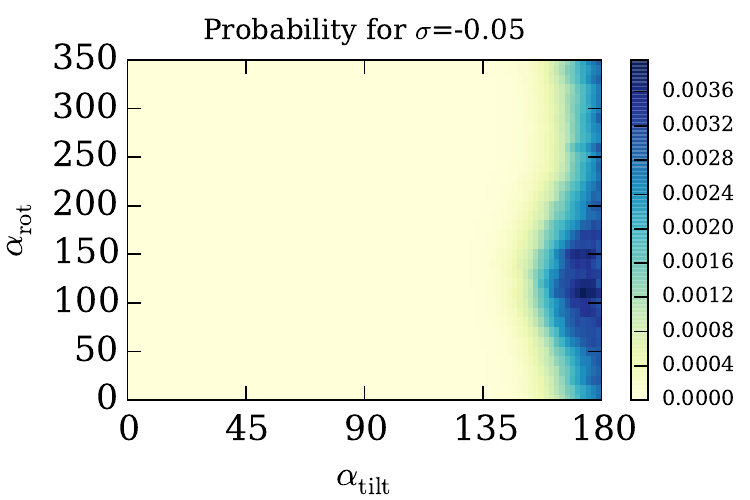} \label{fig:1PGB_2D_sig-005}}
   \caption{Orientation probability distribution of protein \gb. Figures \ref{fig:1PGB_cos} and \ref{fig:1PGB_alpha} are the probability with respect to the tilt angle and its cosine, respectively. Figures \ref{fig:1PGB_2D_sig005} and \ref{fig:1PGB_2D_sig-005} are the probability distributions with respect to both the tilt and rotation angles. Data sets, figure files and running/plotting scripts are available under \ccby.\cite{CooperBarba2015-share1348804}}
   \label{fig:1PGB_probability}
\end{figure*}

\begin{figure*}
   \centering
   \subfloat[Negative surface charge ($\alpha_\text{tilt}=172^\circ$, $\alpha_\text{rot}=110^\circ$)]{\includegraphics[width=0.48\textwidth]{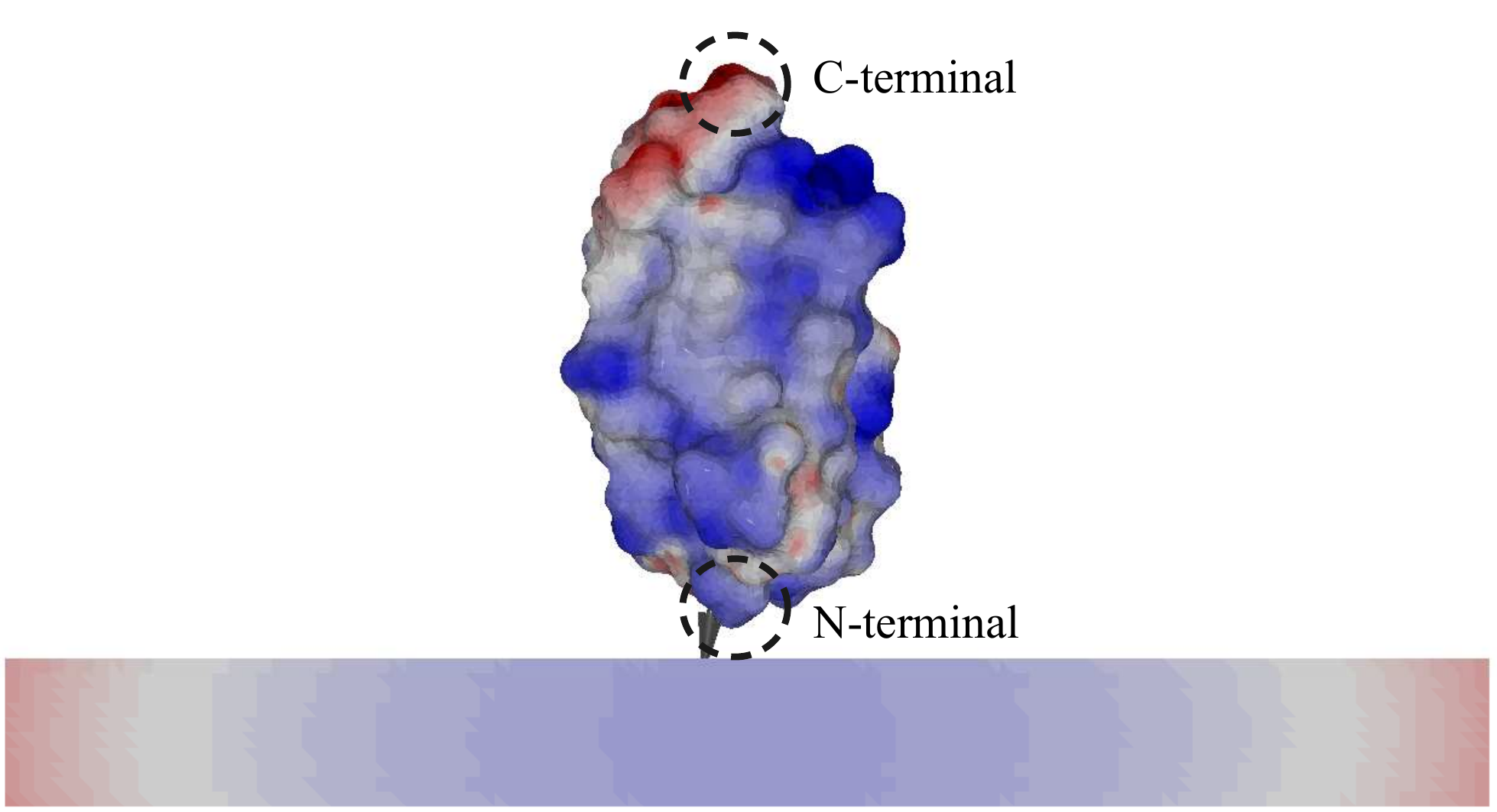} \label{fig:phi_sig-0.05}} 
   \subfloat[Positive surface charge ($\alpha_\text{tilt}=8^\circ$, $\alpha_\text{rot}=150^\circ$)]{\includegraphics[width=0.48\textwidth]{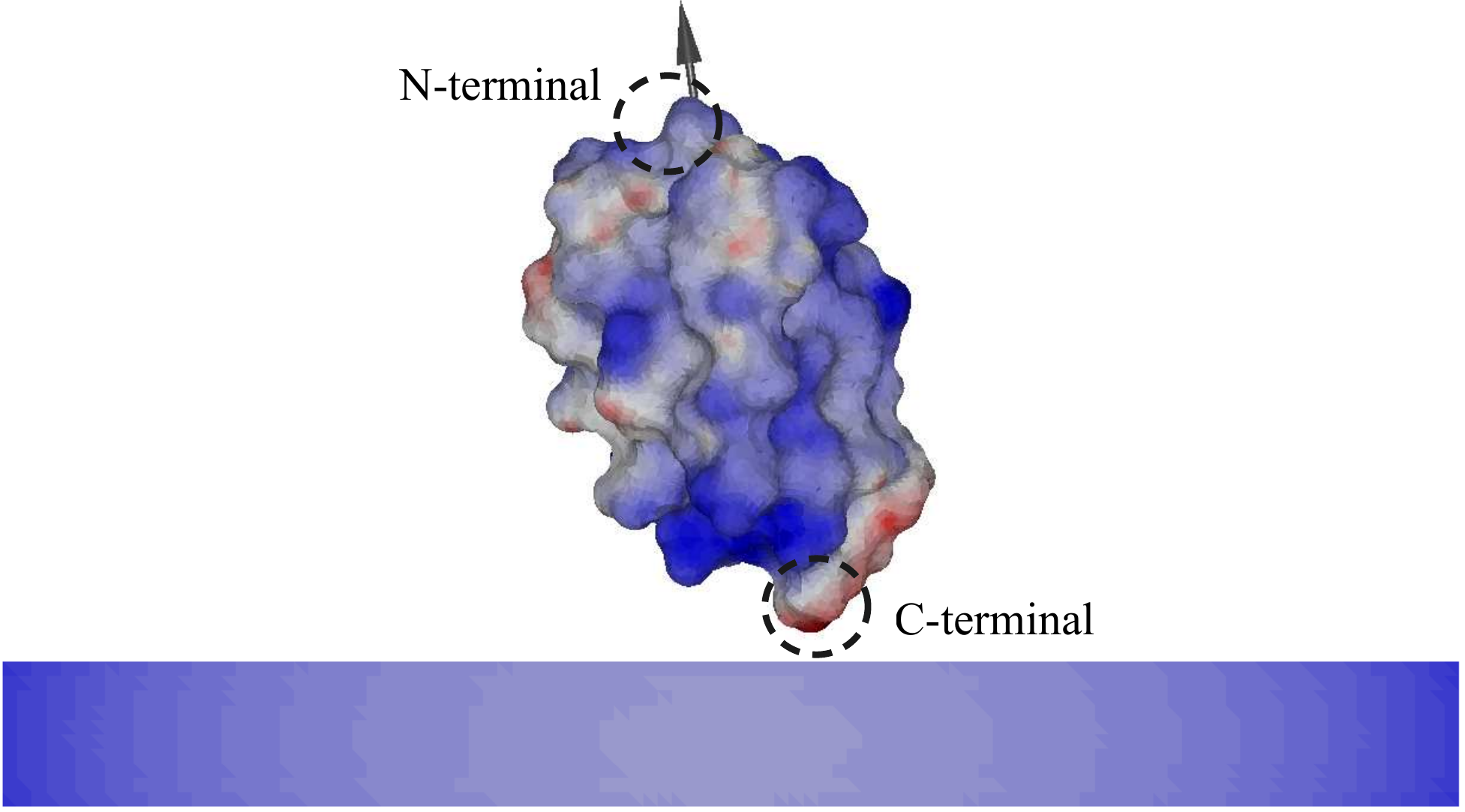} \label{fig:phi_sig0.05}} 
   \caption{Electrostatic potential of protein \gb for the preferred orientations according to Figure \ref{fig:1PGB_probability}. Black arrow indicates direction of dipole-moment vector.}
   \label{fig:field}
\end{figure*}

Using total free energy as the input, the integrals of Equation \eqref{eq:prob_angle} can be computed by means of the trapezoidal rule. Figure \ref{fig:1PGB_probability} presents the probability of the protein orientation in terms of $\cos(\alpha_{\text{tilt}})$, in intervals of $\Delta \cos(\alpha_{\text{tilt}}) = 0.005$ (Fig.~\ref{fig:1PGB_cos}) and $\Delta \alpha_{\text{tilt}}$=2$^{\circ}$ (Fig.~\ref{fig:1PGB_alpha}). Table \ref{table:avg} presents the average orientation $<\cos(\alpha_{\text{tilt}})>$ for the surface having either positive or negative charge density, and Figure \ref{fig:field} shows the electrostatic potential for the preferred orientation in each case. 

\begin{table}[h]
   \caption{\label{table:avg}Average orientation.} 
    \begin{tabular}{c c}
	\hline
	\multicolumn{2}{c} {$<\cos(\alpha_{\text{tilt}})>$} \\
	Negative & Positive \\
	\hline
	$-0.968$ & $0.963$ \\	
	\hline
    \end{tabular}
\end{table}

\subsection{Second case: immunoglobulin G} \label{sec:IGT}

We computed the electrostatic field of immunoglobulin G---a protein widely used in biosensors---interacting with a 250\AA$\times$250\AA$\times$10\AA\ block, varying the conditions of surface charge and salt concentration. The protein was centered with respect to a  250\AA$\times$250\AA\ face, at a distance 5\AA\ above it. 
In fabrication, antibodies are usually immobilized on the biosensor surface via a cross-linker molecule, which we model here by increasing the distance from the surface.
As before, the solvent had relative permittivity of 80 and the protein of 4.

\medskip

 \paragraph*{Grid-convergence study for immunoglobulin G---}

\begin{figure}[h] %  figure placement: here, top, bottom, or page
   \centering
   \includegraphics[width=0.45\textwidth]{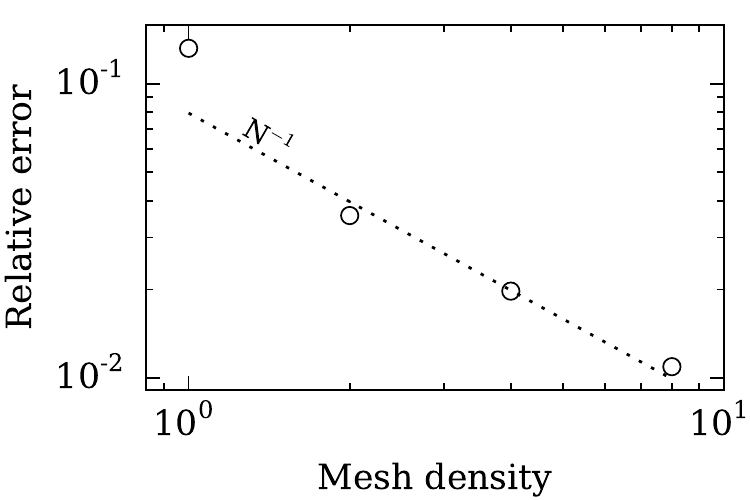} 
   \caption{Grid-convergence study of the solvation plus surface energy for immunoglobulin G interacting with a surface with charge density of 0.05C/m$^2$. Data sets, figure files and plotting scripts are available under \ccby.\cite{CooperBarba2015-share1348801}}
   \label{fig:1IGT_convergence}
\end{figure}

Since this was the first time we did calculations on immunoglobulin G, we carried out a grid-convergence study to make sure the geometry was well resolved and to find adequate values of the simulation parameters for sampling different orientations. The error plotted in Figure \ref{fig:1IGT_convergence} is the relative difference between the energy obtained using \pygbe with each mesh density and the estimated exact value computed with Richardson extrapolation.

In this case, we computed the solvation energy and surface energy of a system consisting of a surface with charge density 0.05C/m$^2$ and a protein with $\alpha_{\text{tilt}} = 31^{\circ}$ and $\alpha_{\text{rot}} = 130^{\circ}$. Using the results from runs with a mesh density of 2, 4, and 8 elements per square Angstrom, we added the solvation and surface energies, and used Richardson extrapolation to obtain a value of $-2792.22$[kcal/mol], and an \emph{observed order of convergence} of 0.85. This is our reference to calculate the errors in Figure \ref{fig:1IGT_convergence}. There is a slight deviation from the expected value of the observed order of convergence (1.0), which we attribute to the non-uniform mesh generated by \msms. Even though the mesh density is on average doubled for each run, there is no guarantee that the refinement is homogeneous throughout the whole molecular surface. The numerical parameters are presented in Table \ref{table:params4}.

\begin{table}[h]
  %\centering
   %\fontfamily{ppl}\selectfont
   \caption{\label{table:params4}Numerical parameters used in the grid-convergence study with immunoglobulin G. } 
    \begin{tabular}{c c c c c c c}
	\hline%\toprule
	\multicolumn{3}{l} {\# Gauss points:} & \multicolumn{3}{l}{Treecode:} & \gmres:\\
	\footnotesize{in-element} & \footnotesize{close-by} & \footnotesize{far-away} & $N_{\text{crit}}$ & $P$ &  $\theta$  & tol.\\
	\hline%\midrule
	9 per side & 19 & 1  &  1000 & 6 & 0.5  & $10^{-5}$\\	
	\hline%\bottomrule
    \end{tabular}
\end{table}

\medskip 

 \paragraph*{Probing orientation of immunoglobulin G---}

We sampled the total free energy every $\Delta \alpha_{\text{tilt}} = 4^\circ$ of tilt angle and $\Delta \alpha_{\text{rot}} =20^\circ$ of rotation angle, resulting in a total of 810 runs.  The surface meshes had 2 triangles per square Angstrom throughout. Numerical parameters are presented in Table \ref{table:params5}.

\begin{table}[h]
  %\centering
   %\fontfamily{ppl}\selectfont
   \caption{\label{table:params5}Numerical parameters used in the runs probing orientation of immunoglobulin G. } 
    \begin{tabular}{c c c c c c c}
	\hline%\toprule
	\multicolumn{3}{l} {\# Gauss points:} & \multicolumn{3}{l}{Treecode:} & \gmres:\\
	\footnotesize{in-element} & \footnotesize{close-by} & \footnotesize{far-away} & $N_{\text{crit}}$ & $P$ &  $\theta$  & tol.\\
	\hline%\midrule
	9 per side & 19 & 1  &  300 & 2 & 0.5  & $10^{-4}$\\	
	\hline%\bottomrule
    \end{tabular}
\end{table}

\begin{figure*} 
   \centering
   \subfloat[Probability for $\sigma=-0.05$C/m$^2$, $\kappa=0.125$\AA$^{-1}$]{\includegraphics[width=0.4\textwidth]{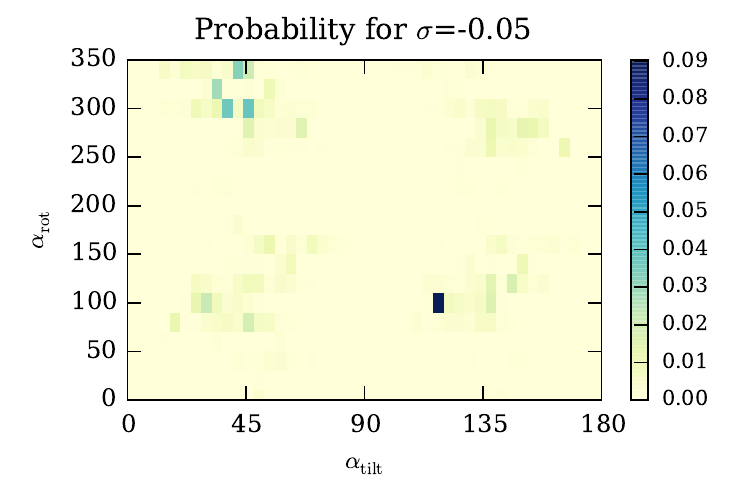} \label{fig:1IGT_2D_sig-005}}
   \subfloat[x-y plane view for $\alpha_{\text{tilt}} = 116^{\circ}$ and $\alpha_{\text{rot}} = 100^{\circ}$]{\includegraphics[width=0.4\textwidth]{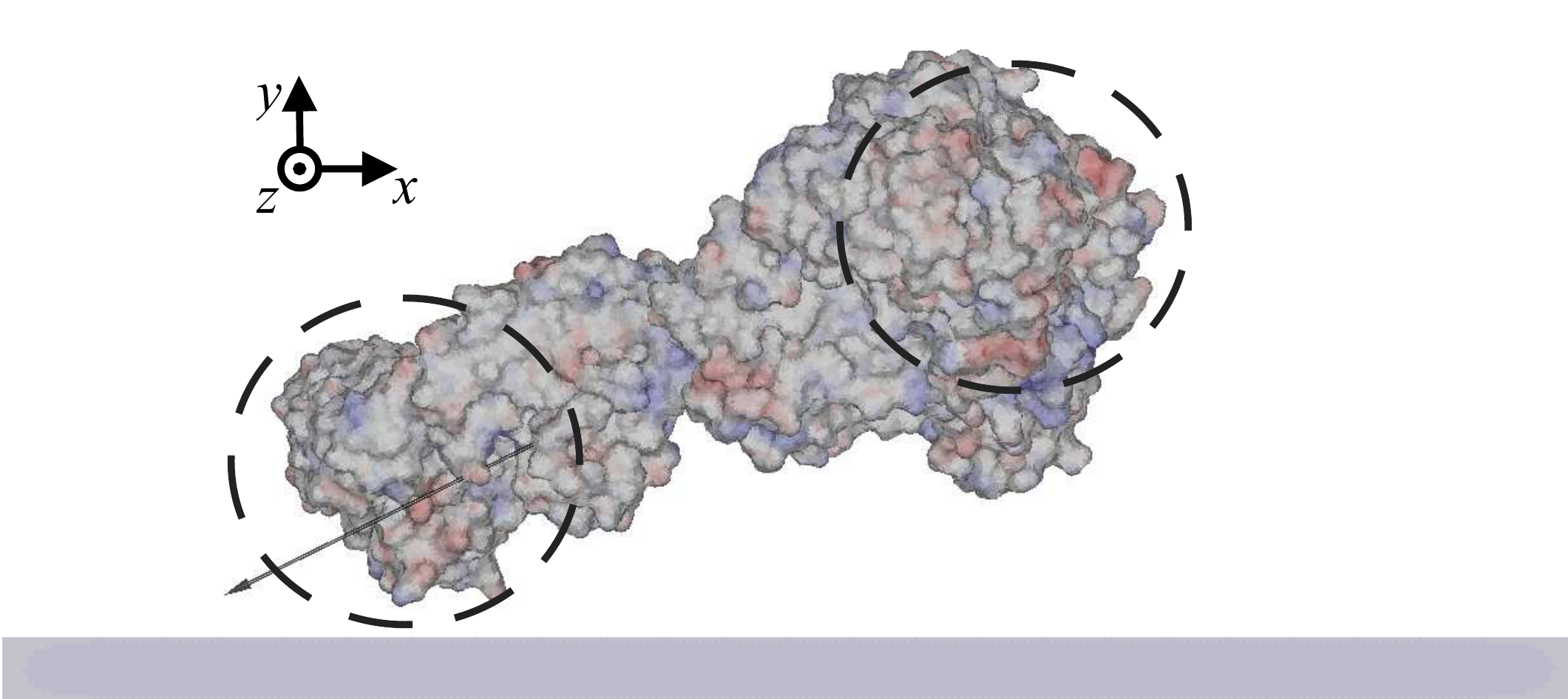} \label{fig:1IGT_3D_sig-005_kap0125_til116-rot100}}\\
   \subfloat[Probability for $\sigma=-0.1$C/m$^2$, $\kappa=0.125$\AA$^{-1}$]{\includegraphics[width=0.4\textwidth]{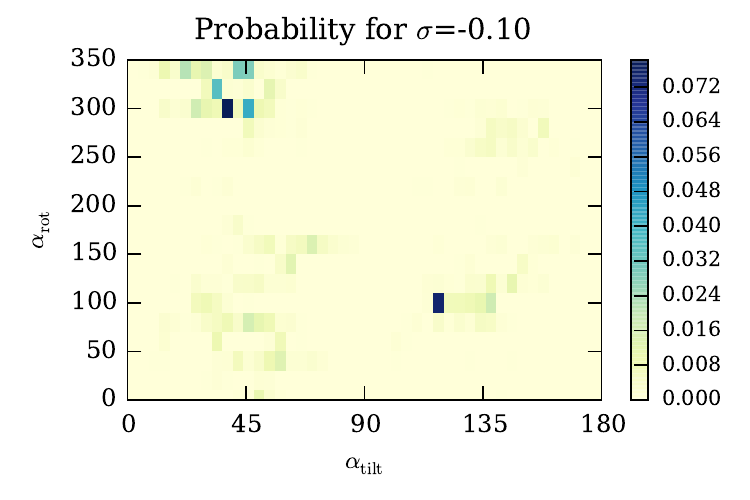} \label{fig:1IGT_2D_sig-020_kappa01250}}
   \subfloat[y-z plane view for $\alpha_{\text{tilt}} = 36^{\circ}$ and $\alpha_{\text{rot}} = 300^{\circ}$]{\includegraphics[width=0.4\textwidth]{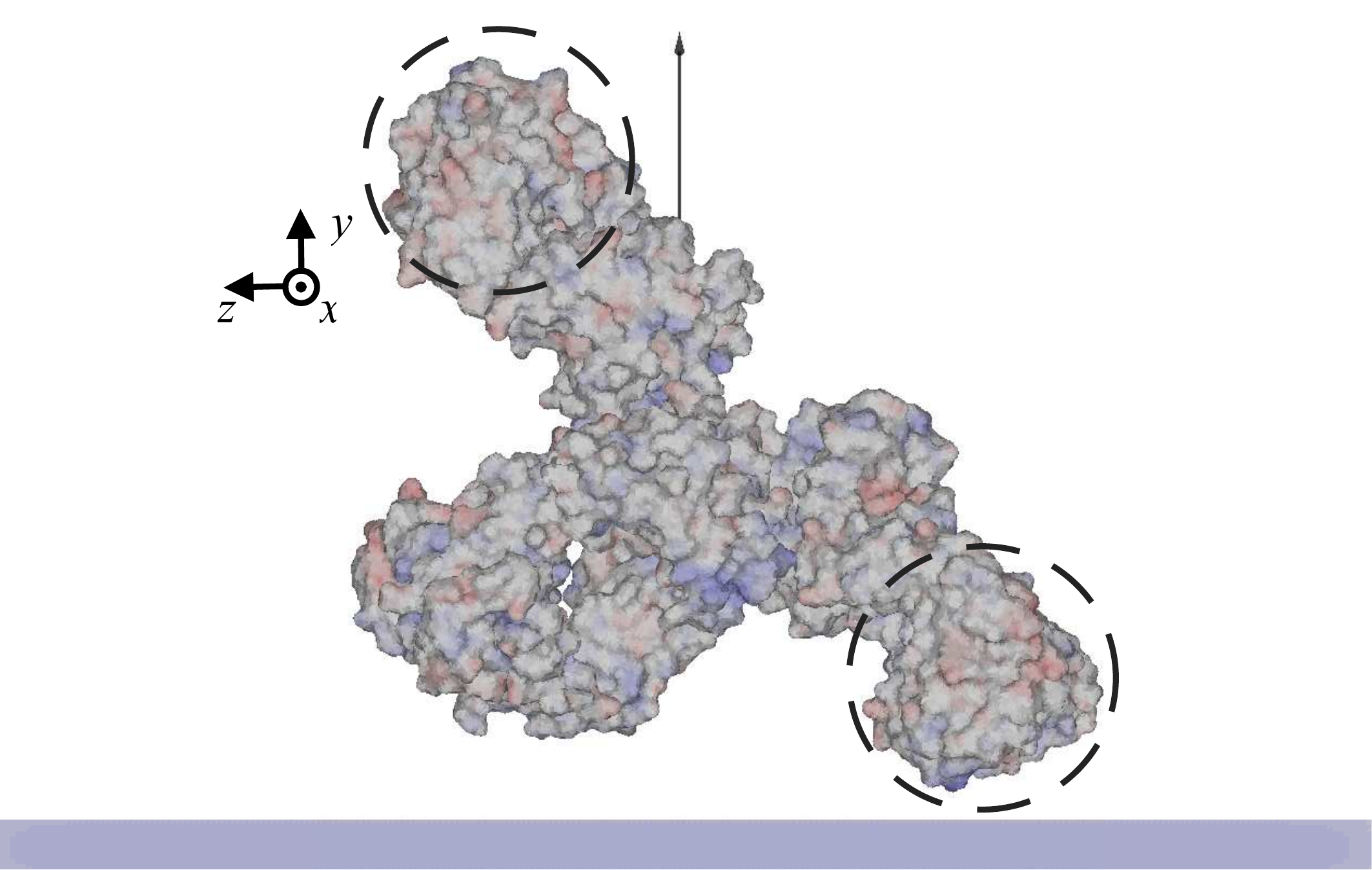} \label{fig:1IGT_3D_sig-02_kap0125_til056-rot040}}\\
   \subfloat[Probability for $\sigma=-0.05$C/m$^2$, $\kappa=0.0625$\AA$^{-1}$]{\includegraphics[width=0.4\textwidth]{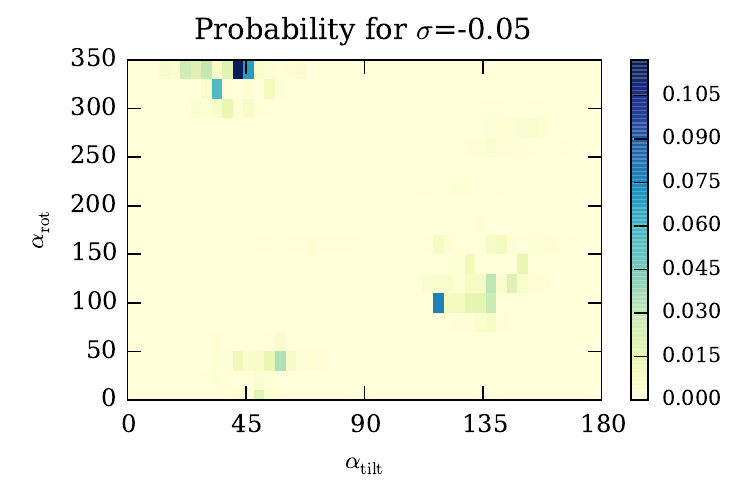} \label{fig:1IGT_2D_sig-005_kappa003125}}
   \subfloat[x-y plane view for $\alpha_{\text{tilt}} = 40^{\circ}$ and $\alpha_{\text{rot}} = 340^{\circ}$]{\includegraphics[width=0.4\textwidth]{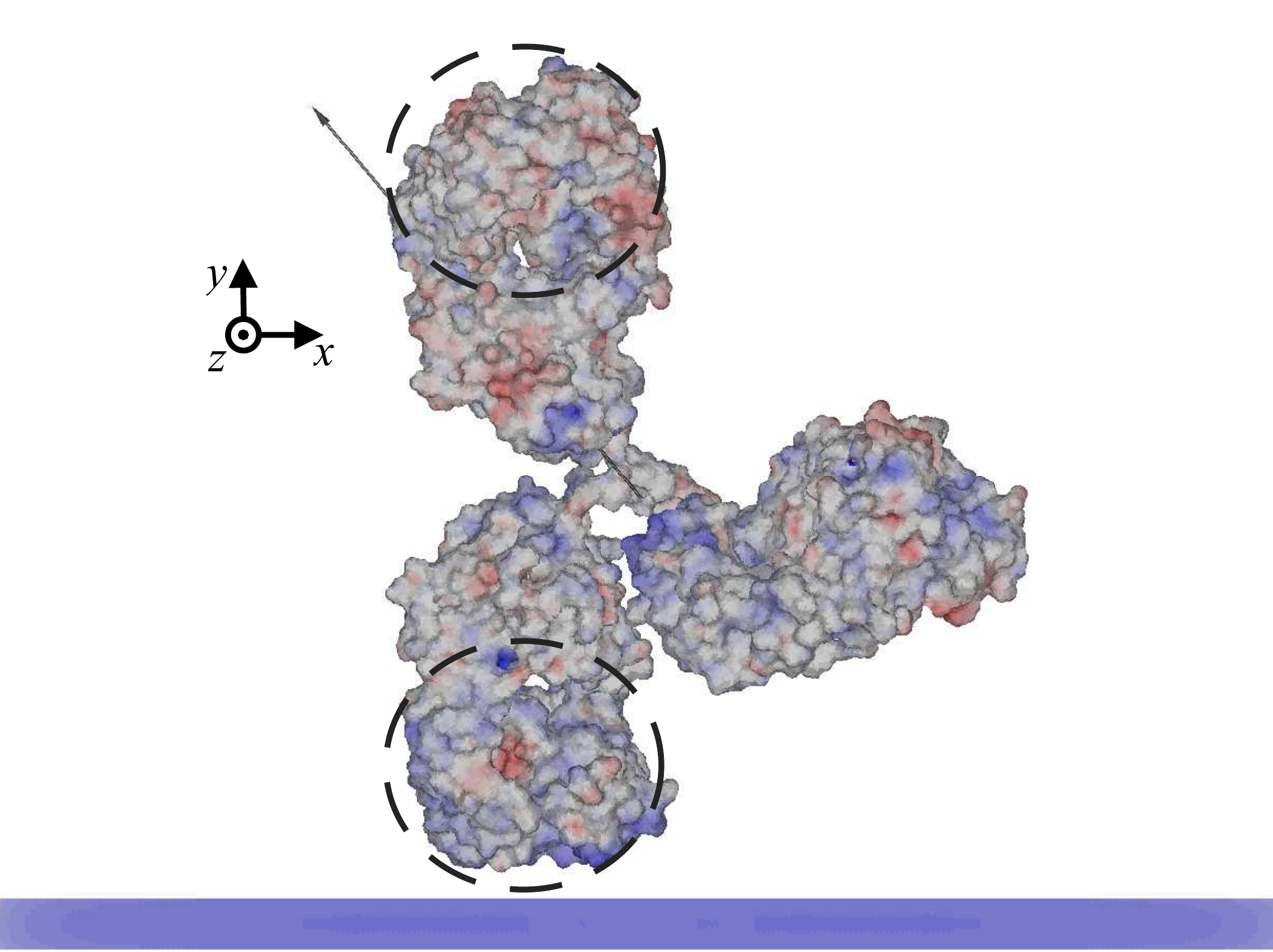} \label{fig:1IGT_3D_sig-005_kap003125_til116-rot160}}\\
   \subfloat[Probability for $\sigma=-0.1$C/m$^2$, $\kappa=0.0625$\AA$^{-1}$]{\includegraphics[width=0.4\textwidth]{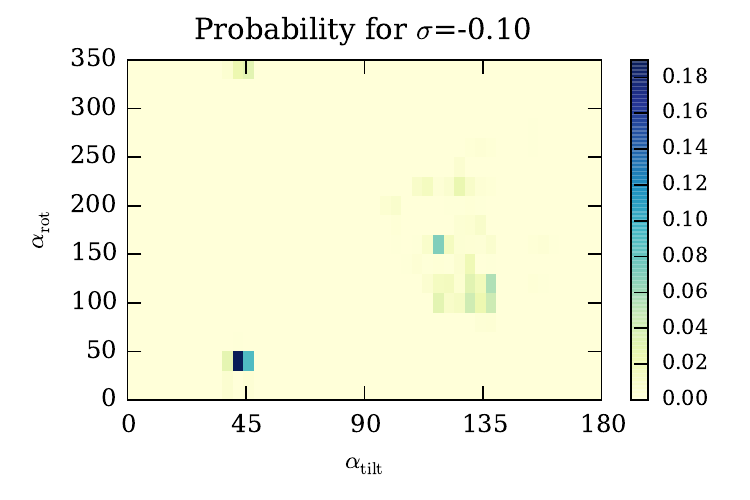} \label{fig:1IGT_2D_sig-020_kappa003125}}
   \subfloat[x-y plane view for $\alpha_{\text{tilt}} = 40^{\circ}$ and $\alpha_{\text{rot}} = 40^{\circ}$]{\includegraphics[width=0.4\textwidth]{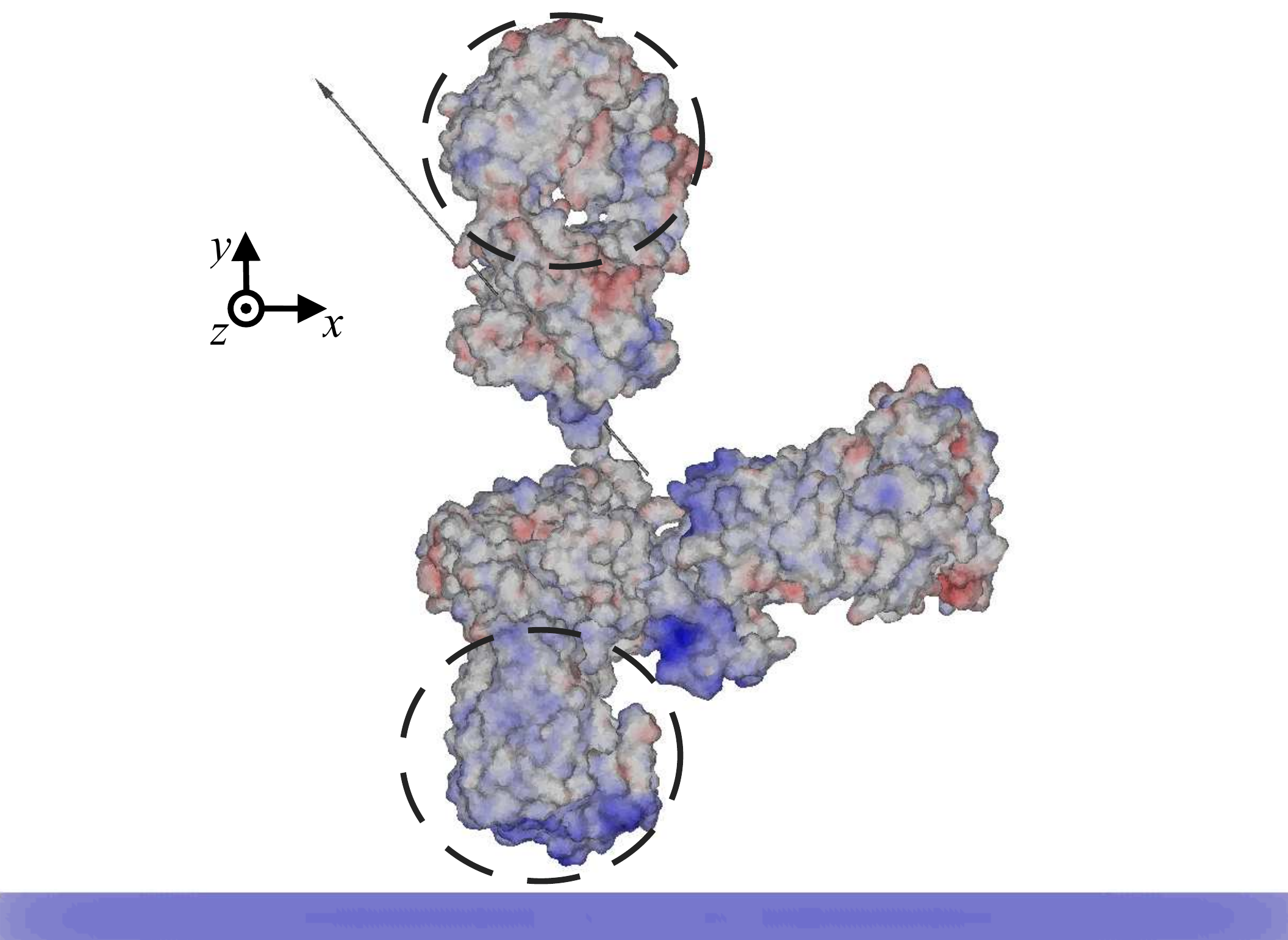} \label{fig:1IGT_3D_sig-02_kap003125_til124-rot140}}
   \caption{Orientation probability distribution and surface potential of the preferred orientation for immunoglobulin G near a negative surface charge. The black arrow indicates the direction of the dipole moment, and the circles enclose the Fab fragments. Data sets, figure files and plotting scripts available under \ccby.\cite{CooperBarba2015-share1348802}}
   \label{fig:1IGT_negcharge}
\end{figure*}

\begin{figure*}
   \centering
   \subfloat[Probability for $\sigma$=0.05C/m$^2$ and $\kappa$=0.125\AA$^{-1}$]{\includegraphics[width=0.4\textwidth]{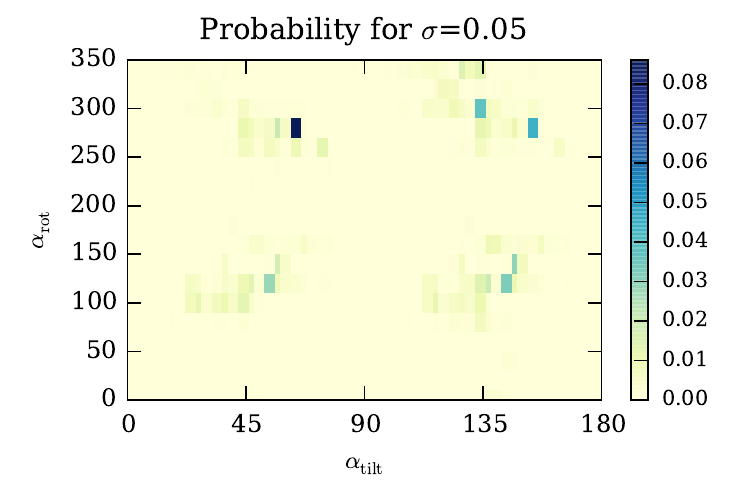} \label{fig:1IGT_2D_sig005}}
   \subfloat[x-y plane view for $\alpha_{\text{tilt}} = 64^{\circ}$ and $\alpha_{\text{rot}} = 280^{\circ}$]{\includegraphics[width=0.4\textwidth]{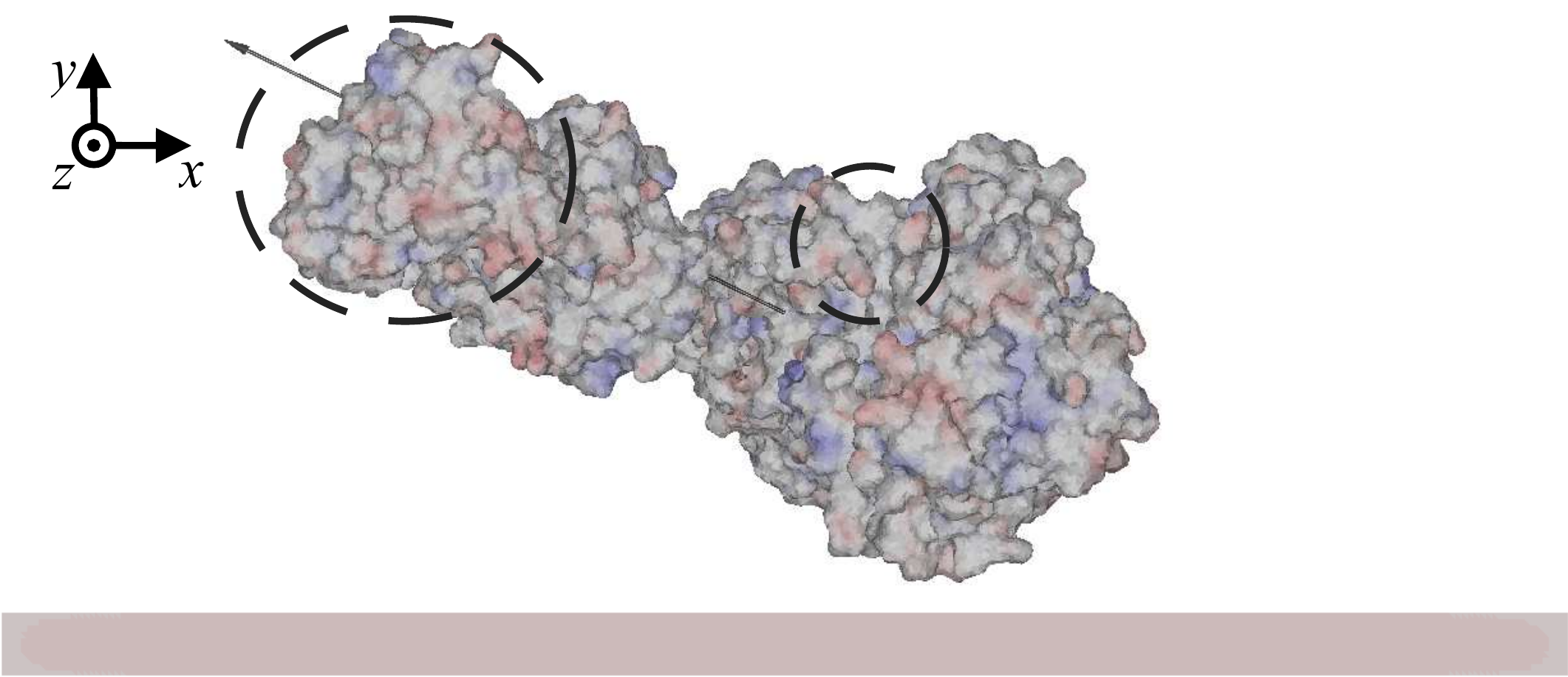} \label{fig:1IGT_3D_sig005_kap0125_til064-rot280}}\\
   \subfloat[Probability for $\sigma$=0.1C/m$^2$ and $\kappa$=0.125\AA$^{-1}$]{\includegraphics[width=0.4\textwidth]{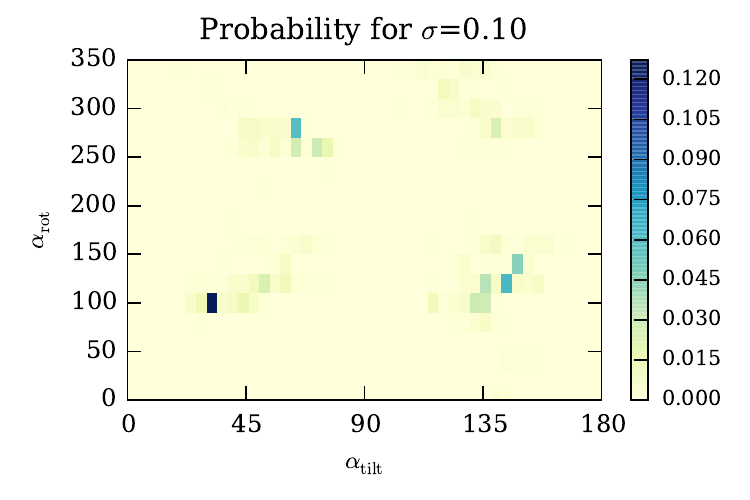} \label{fig:1IGT_2D_sig020_kappa01250}}
   \subfloat[y-z plane view for $\alpha_{\text{tilt}} = 32^{\circ}$ and $\alpha_{\text{rot}} = 100^{\circ}$]{\includegraphics[width=0.4\textwidth]{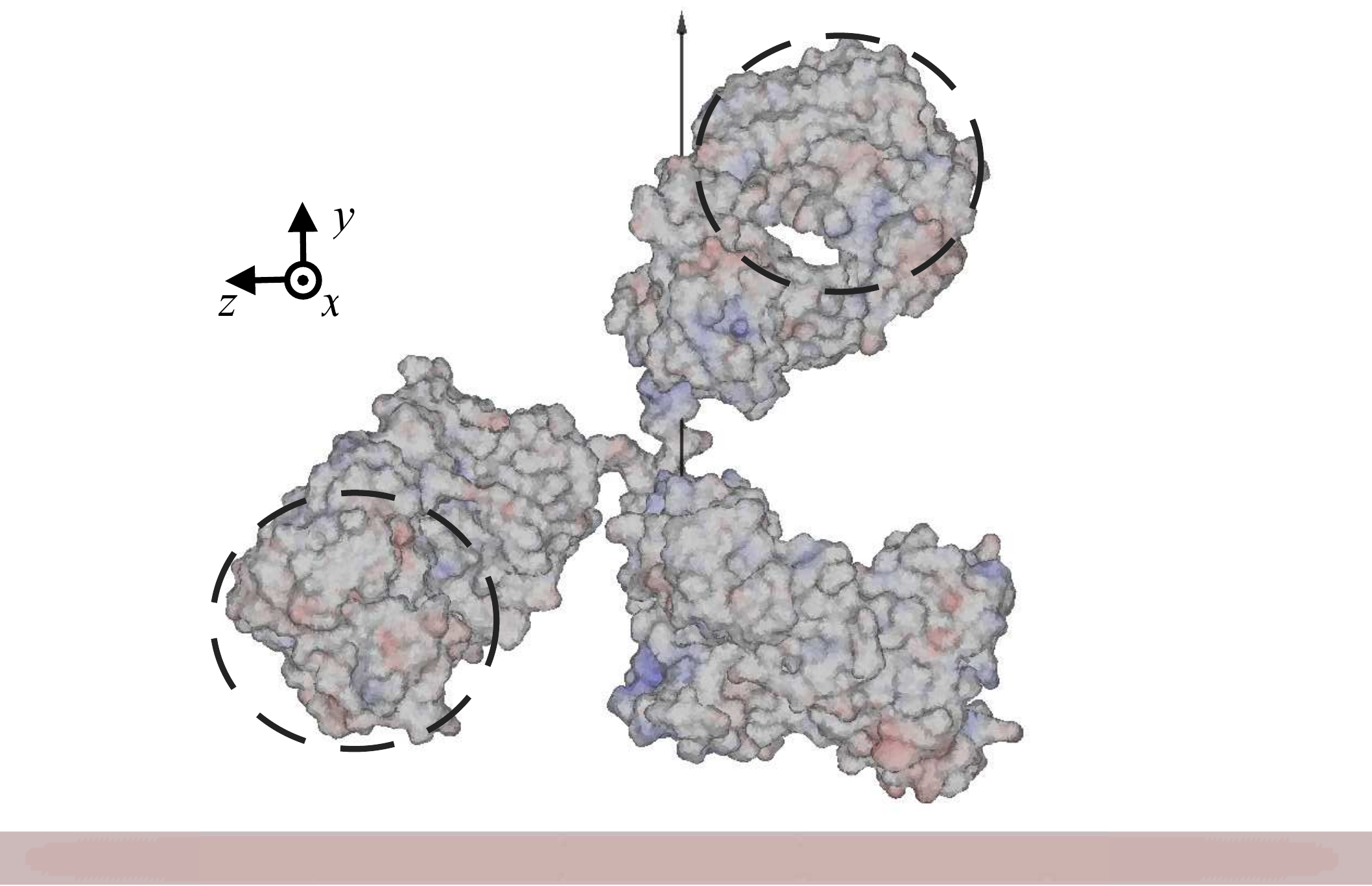} \label{fig:1IGT_3D_sig02_kap0125_til064-rot260}}\\
   \subfloat[Probability for $\sigma$=0.05C/m$^2$ and $\kappa$=0.0625\AA$^{-1}$]{\includegraphics[width=0.4\textwidth]{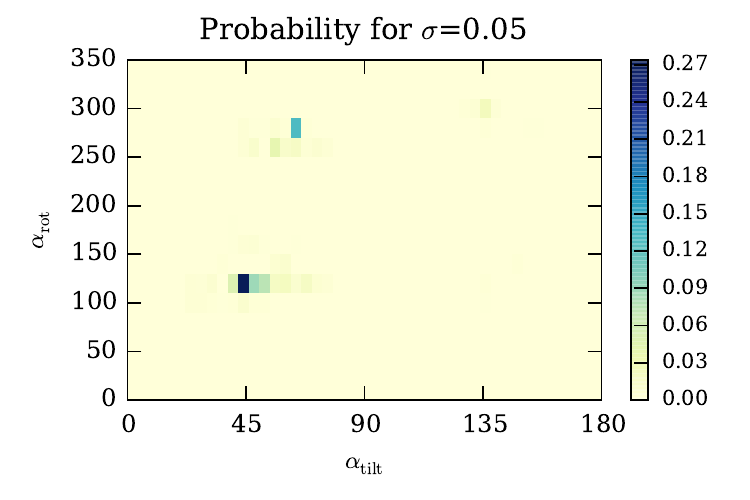} \label{fig:1IGT_2D_sig005_kappa003125}}
   \subfloat[y-z plane view for $\alpha_{\text{tilt}} = 44^{\circ}$ and $\alpha_{\text{rot}} = 120^{\circ}$]{\includegraphics[width=0.4\textwidth]{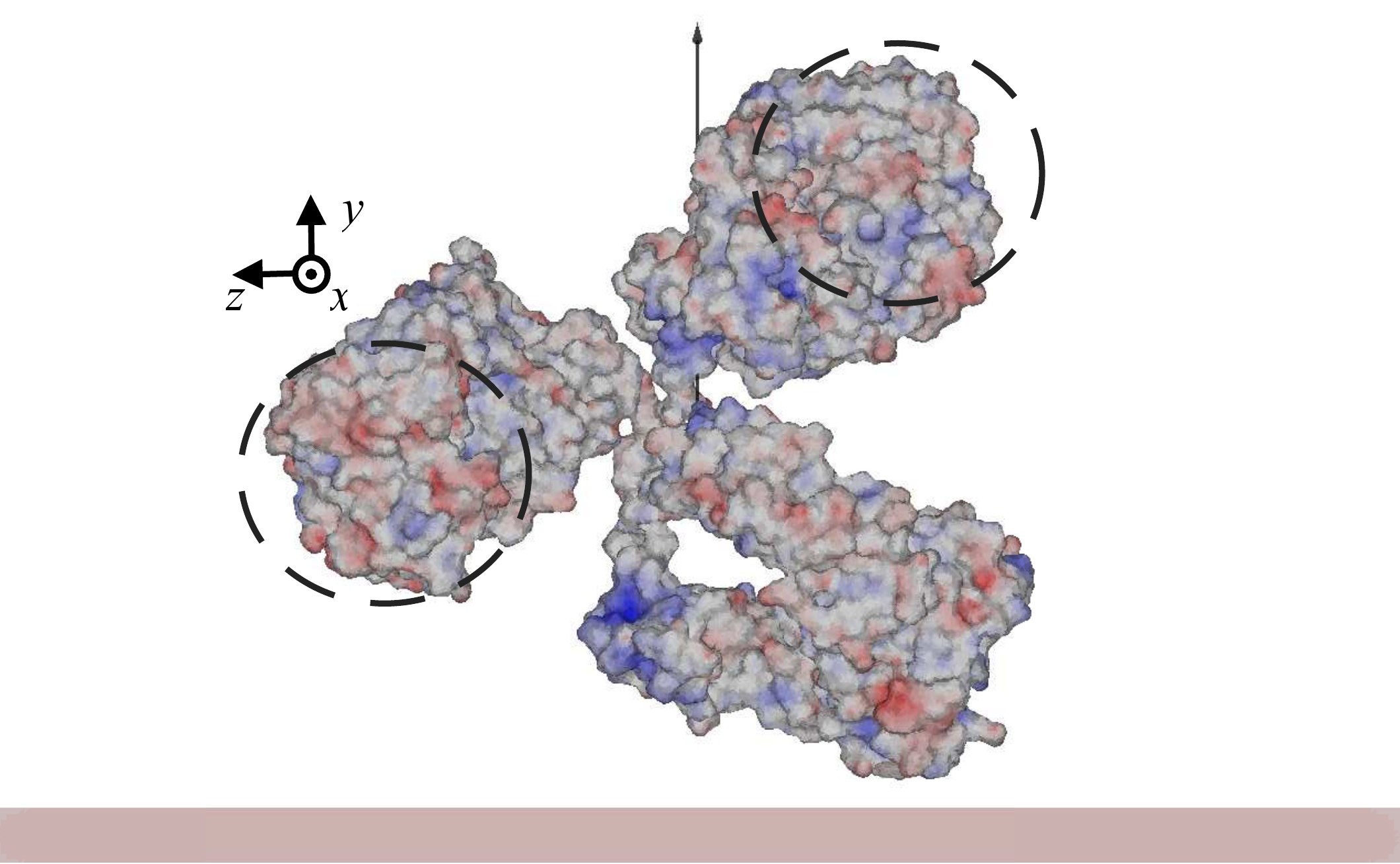} \label{fig:1IGT_3D_sig005_kap003125_til044-rot120}}\\
   \subfloat[Probability for $\sigma$=0.1C/m$^2$ and $\kappa$=0.0625\AA$^{-1}$]{\includegraphics[width=0.4\textwidth]{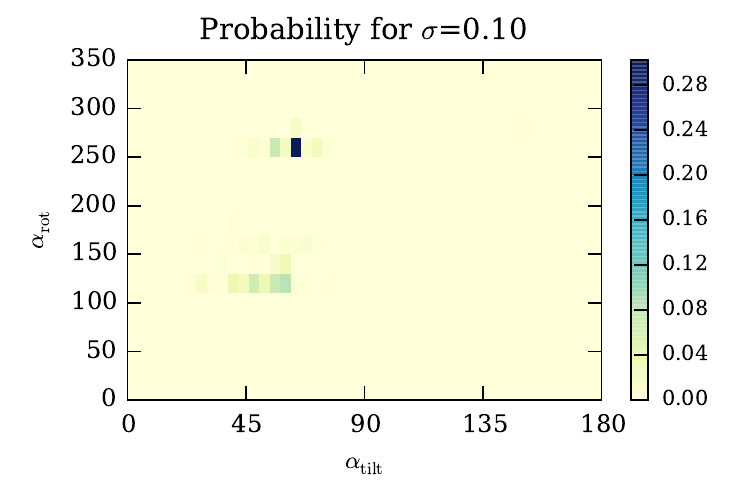} \label{fig:1IGT_2D_sig020_kappa003125}}
   \subfloat[x-y plane view for $\alpha_{\text{tilt}} = 64^{\circ}$ and $\alpha_{\text{rot}} = 260^{\circ}$]{\includegraphics[width=0.4\textwidth]{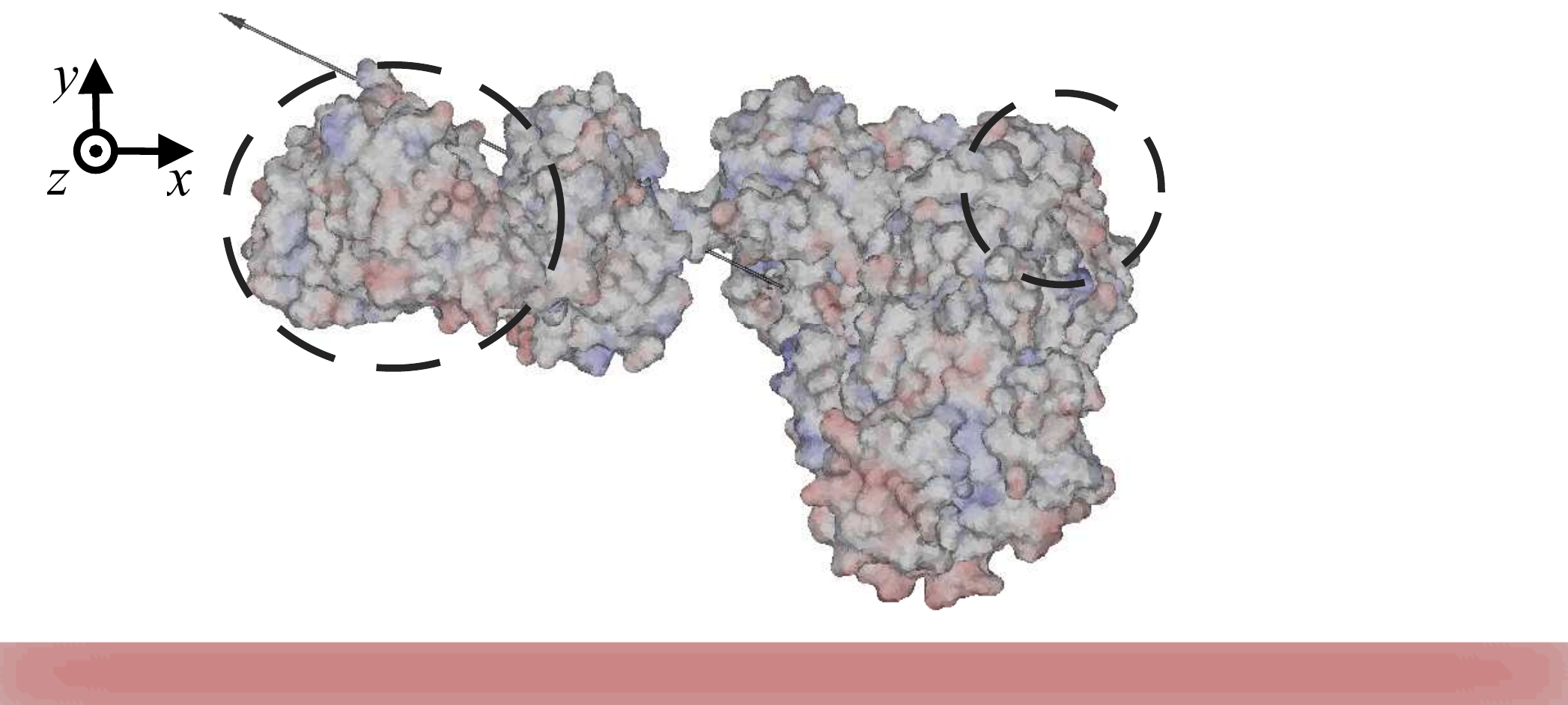} \label{fig:1IGT_3D_sig02_kap003125_til076-rot160}}
   \caption{Orientation probability distribution and surface potential of the preferred orientation for immunoglobulin G near a positive surface charge. The black arrow indicates the direction of the dipole moment, and the circles enclose the Fab fragments. Data sets, figure files and plotting scripts available under \ccby.\cite{CooperBarba2015-share1348802}}
   \label{fig:1IGT_poscharge}
\end{figure*}

With the computed total free energy, we obtained the probability of each orientation using Equation \eqref{eq:prob_angle} and the trapezoidal rule.  
We sampled all combinations with surface charges of $\sigma=\pm$0.05C/m$^2$ and $\sigma = \pm$ 0.1C/m$^2$ and salt concentrations of 145mM ($\kappa$ = 0.125 \AA$^{-1}$) and 37mM ($\kappa$ = 0.0625 \AA$^{-1}$). For each of these cases, Figures \ref{fig:1IGT_negcharge}  and \ref{fig:1IGT_poscharge} show a color plot of the probability distribution with respect to the tilt and rotation angles, and a 3D plot of the preferred orientation, where the solvent-excluded surface is colored by the electrostatic potential.

% FIGURE 8

\subsection{Reproducibility and data management}
We have a consistent reproducibility practice that includes releasing code and data associated with a publication. The \pygbe code was released at the time of submitting our previous publication,\cite{CooperBardhanBarba2013} under an MIT open-source license, and we maintain a version-control repository. As with our previous paper, we also release with this work all of the data needed to run the numerical experiments reported here, including running scripts and post-processing code in Python for producing the figures. To support our open-science goals, we prepared such a \emph{``reproducibility package''} for each of the results presented in Figures \ref{fig:1PGB_probability}, \ref{fig:1IGT_convergence}, and the probability plots in Figures \ref{fig:1IGT_negcharge} and \ref{fig:1IGT_poscharge}. The included running scripts invoke the \pygbe code with the correct input data and meshes (also included), and post-process the results to give the final figure, all with just one command. Please see the respective captions for a reference to the reproducibility packages, hosted on the \textbf{figshare} repository.

\section{Discussion} \label{sec:discussion}
%!TEX root = CooperBarba-orientation.tex

\subsection{First case: protein G\,B1\,D4$^\prime$} \label{sec:disc_1PGB}

The orientation of protein \gb near charged surfaces was studied using a combined Monte Carlo and molecular dynamics method by Liu and co-workers\cite{LiuLiaoZhou2013} and experimentally by Baio and co-workers.\cite{BaioWeidnerBaughGambleStaytonCastner2012} The availability of these published results was a motivation to use this protein for a first test, to compare with the results obtained with our model. 

The results presented in Figure \ref{fig:1PGB_probability} show that for the most likely orientations, the dipole-moment vector is aligned with the vector normal to the interacting surface. This indicates that the dipole moment is the dominant effect that determines the protein's orientation, over local protein-surface  interactions. This is the expected result, since protein \gb is a relatively small biomolecule. 

Moreover, Figure \ref{fig:1PGB_probability} reveals that protein \gb behaves like a point dipole, as the most likely orientations shift 180$^\circ$ when the sign of the surface charge is flipped. This is also explained by the dipole moment dominating the orientation.
In fact, we repeated this whole set of calculations but placing protein \gb at a greater distance, 5\AA~ away from the surface, and the results did not vary.

The dipolar behavior described by our results agrees with the experiments done by Baio and co-workers, \cite{BaioWeidnerBaughGambleStaytonCastner2012} in which they observed opposite orientations of protein \gb adsorbed on NH$_3^+$ and COO$^-$ self-assembled monolayers. With positively charged surfaces, most of the proteins oriented with the N-terminal of the protein pointing away from the surface, while for negatively charged surfaces the opposite occurred, with the C-terminal pointing away from the surface. This agrees with our results in Figure \ref{fig:1PGB_probability} (the dipole moment vector of protein \gb points from the C-terminal to the N-terminal).

Liu and co-workers \cite{LiuLiaoZhou2013} used a combined Monte Carlo and molecular dynamics method to obtain $<\cos(\alpha_{\text{tilt}})>=0.95$ for $\sigma = 0.05$C/m$^2$, and $<\cos(\alpha_{\text{tilt}})>=-0.85\pm0.05$ for $\sigma = -0.05$C/m$^2$, which agrees well with our results in Table \ref{table:avg}. Note that MD simulations consider van der Waals interactions and conformational changes of the protein, whereas these are not considered in our approach, explaining the slight differences in $<\cos(\alpha_{\text{tilt}})>$.
However, as noted by other researchers,\cite{ZhouChenJiang2003,BaioWeidnerBaughGambleStaytonCastner2012,LiuLiaoZhou2013} electrostatic effects often dominate protein-surface interactions and drive orientation during adsorption, while van der Waals effects play a role only in cases of very low surface charge. For example, in Ref.~\onlinecite{ZhouChenJiang2003}, van der Waals effects were of consequence in a setup with surface charge of 0.006C/m$^{2}$ and high ionic strength, leading to weak electrostatics. In a biosensor-fabrication scenario, this would only be the case with low-quality \sam s.

The results with protein \gb mean that an electrostatic solver with implicit solvent using the Poisson-Boltzmann equation is capable of capturing the driving mechanism of physical adsorption and orientation of the adsorbed molecule, at least in cases where the molecule's dipole moment is dominating the orientation. This is important because protein adsorption, being a free energy-driven process, is difficult to study experimentally\cite{MijajlovicETal2013} and thus simulations offer a promising alternative. Full atomistic molecular dynamics, however, demands large amounts of computing effort, and the possibility of using an electrostatics solver may extend the range of systems that can be investigated.

 \subsection{Second case: immunoglobulin G}

With our numerical model already verified using an analytical solution for spherical geometry\cite{CooperBarba2015a} and the successful results for protein orientation of a small protein near a charged surface (Section \ref{sec:disc_1PGB}), we proceeded to study the effect of surface charge and salt concentration on the orientation of the antibody immunoglobulin G. Antibodies are widely used in biosensors as ligand molecules, due to their affinity and specificity with the target molecule (antigen), and it is vitally important that they are adsorbed on the sensor with the antigen-binding Ig fragment (Fab) pointing away from the sensor, into the oncoming flow containing the antigens (known as ``end-on'' or ``tail-on'' orientation).
Early experimental studies found that antigen/antibody ratio was especially low on negatively charged surfaces,\cite{BuijsETal1997} leading to the notion that protein orientation was affected to leading order by charge. 
One subsequent study\cite{ChenLiuZhouJiang2003} investigated the orientations of two iso-types of immunoglobulin G---\ig 1, corresponding to \pdb\ structure {\small 1IGY}, and \ig 2a, corresponding to \pdb\ {\small 1IGT}---adsorbed on positive and negatively charged surfaces. 
As an indirect method of probing antibody orientation, the researchers obtained adsorbed amounts and antigen/antibody ratios by means of surface-plasmon resonance experiments (e.g., a higher antigen/antibody ratio would indicate that more active sites are accessible and more antibodies are in a favorable orientation). 
The finding was that \ig 1 mainly had a ``head-on'' (unfavorable) orientation on the negatively charged surfaces and a mix of ``tail-on'' (most favorable) and ``side-on'' orientations on the positively charged surfaces. 
\ig 2a, on the other hand, had many orientations on both surfaces with positive and negative charge, leading to the conclusion that \ig 2a is harder to control using electrostatic effects.
Results consistent with these were obtained by Zhou and co-workers\cite{ZhouChenJiang2003} using a united-residue model: a coarse-grained model where each amino-acid is treated as a sphere. They find that \ig 1 will have the favorable ``end-on'' orientation on positive surfaces, as long as the charge density was large enough (0.018C/m$^{2}$, in their case) and the ionic strength was low. But \ig 2a  did not show a clear preferred orientation at the conditions they looked at; the authors attribute this to the weaker dipole moment of this iso-type.
 
We investigated the orientation of \ig 2a, which other studies found harder to orient favorably on a biosensor surface, and used two values of the surface charge ($\sigma=0.05$ and $0.1$C/m$^{2}$) and two values of salt concentration ($\kappa=0.125$ and $0.0625$\AA$^{-1}$), in each case varying two-fold.
 Figures \ref{fig:1IGT_negcharge} and \ref{fig:1IGT_poscharge} present the probability distribution of \ig 2a for many orientations (given by $\alpha_\text{tilt}$ and $\alpha_\text{rot}$), in each case.
 The following discussion refers to each variation of the parameters and the effect on the preferred orientation of the adsorbed antibody and its probability.

 \medskip
 
 \paragraph*{Effect of surface charge---}
 
The lower value of surface charge here is $\sigma=\pm 0.05$C/m$^2$, the same value used in Ref.~\onlinecite{LiuLiaoZhou2013} to mimic the experiments reported in Ref.~\onlinecite{BaioWeidnerBaughGambleStaytonCastner2012}. 
Figures \ref{fig:1IGT_2D_sig-005} and \ref{fig:1IGT_2D_sig005} show that for the lower value of surface charge with the higher salt concentration ($\kappa=0.125$\AA$^{-1}$), there is no clear preferred orientation, to the point that the highest probability falls under 10\%. 
This means that adsorbing the antibodies under these conditions would result in a wide range of orientations, which would not be favorable for biosensor fabrication.
Moreover, the preferred configurations in figures \ref{fig:1IGT_2D_sig-005} and \ref{fig:1IGT_2D_sig005} show the antibody lying flat on the surface, far from the desired ``tail on'' orientation. 
This observation is consistent with a previous study using a unified-residue model,\cite{ZhouChenJiang2003} where this particular antibody showed many possible orientations. 
The authors of that study attributed this behavior to the weaker dipole moment of this molecule, compared with the variant \ig 1.
 
With the higher value of surface charge, in this case $\sigma=\pm0.1$C/m$^2$, the orientation probability distribution in the case of negative charge improves somewhat, as the antibody is slanted sideways rather than lying down for $\kappa=0.125$\AA$^{-1}$ (at least one antingen-binding fragment is pointing up), and the probability of the preferred orientation is almost doubled for low salt concentration, in a ``side on'' orientation.
For positive surface charge the slanted orientation is similar, however the probability is higher for the preferred orientation in both the low- and high-salt cases.
In the cases with higher salt concentration, Figure \ref{fig:1IGT_2D_sig020_kappa01250} shows a preferred orientation with a higher probability of 12\%, compared to 8\% in Figure \ref{fig:1IGT_2D_sig005}, and the dipole moment rotates towards the normal vector. 
For the lower value of salt concentration (Figure \ref{fig:1IGT_2D_sig020_kappa003125}), this effect is smaller, however it shifts the preferred tilt angle in the opposite direction, from $44^{\circ}$ to $64^{\circ}$. Note that the dipole-moment vector does not point straight through the middle between the two Fab fragments, but in an angle.
This indicates that, in contrast to protein \gb, local interactions dominate over the dipole moment. If the dipole moment were the dominant effect, the dipole-moment vector would tend to align to the surface normal as the surface charge increases.
This argues against the suggestion by other researchers\cite{ChenLiuZhouJiang2003,ZhouChenJiang2003} that the dipole-moment vector is the main determinant of orientation.
 
 \medskip
 
 \paragraph*{Effect of salt concentration---}
 
As the surface charge density was varied two-fold, we also varied the Debye length ($\kappa^{-1}$) two-fold. In terms of salt concentration, it means a 4$\times$ decrease in the amount of salt. The higher value of salt concentration corresponds to 145mM, which is in the physiological salt range.  
 
Like increasing the surface charge, lowering the salt concentration affects the orientation probability distribution. 
 For $\sigma=-0.05$C/m$^2$ (Fig.~\ref{fig:1IGT_2D_sig-005_kappa003125}), the effect is a large shift in the preferred tilt angle, from $\alpha_\text{tilt}=116^\circ$ to $\alpha_\text{tilt}=40^\circ$, with a small change in probability. 
For the positive weaker charge, $\sigma=0.05$C/m$^2$ (Fig.~\ref{fig:1IGT_2D_sig005_kappa003125}), not only does the peak probability increase considerably ($\sim 3\times$), but the preferred tilt shifts from $64^{\circ}$ to $44^{\circ}$.
This orientation is favorable for biosensing applications, as the antigen-binding fragments are pointing away from the surface, in a ``tail on'' orientation.
 For the stronger negative charge, $\sigma=-0.1$C/m$^2$, the probability peak increases  $2.5\times$ for a ``side on'' orientation where one of the antigen-binding fragments is attached to the surface in an unfavorable position (Fig.~\ref{fig:1IGT_2D_sig-020_kappa003125}).
 With positive surface charge, the tilt angle shifts in such a way that the antibody is lying on the surface with a marked probability close to $30\%$ (Fig.~\ref{fig:1IGT_2D_sig020_kappa003125}).
 This orientation is not ideal for biosensors, but it is better than the slanted position as neither of the Fabs are attached to the surface.

From the results in Figures \ref{fig:1IGT_negcharge} and \ref{fig:1IGT_poscharge}, we conclude that the iso-type \ig 2a can in general be better orientated with low salt concentration and high surface charge, as we get more pronounced high-probability regions. 
Moreover, good orientations for biosensors are more likely to occur with positive surface charge (Figure \ref{fig:1IGT_3D_sig005_kap003125_til044-rot120}), since the Fab fragments are pointing up.
Previous studies had shown that the \ig 1 variant could be controlled, but not \ig 2a. 
The advantage of a positive surface charge and a low ionic strength had been suggested by previous studies, but not for this particular variant of immunoglobulin G. Note also that our lower value of salt concentration is 37mM, which is a higher amount of salt than other studies.\cite{BuijsETal1997,ChenLiuZhouJiang2003}

 \medskip
A limitation of this study stems from the application of linearized Poisson-Boltzmann equation.
Bu and co-workers\cite{BuVakninTravesset2006} assessed the accuracy of the Poisson-Boltzmann equation for highly charged surfaces ($\sim 0.4$C/m$^2$), getting good agreement of the model with experiments. 
Rigorously, the linearized Poisson-Boltzmann equation is a valid approximation of the Poisson-Boltzmann equation when the nondimensional potential is smaller than 1 ($\phi q_e/k_BT<<1$).
However, this restriction can be relaxed when calculating solvation energy. 
For example, we ran a calculation using our boundary element code on an isolated \ig 2a immersed in a solvent with 37mM of salt ($\kappa = 0.0625$\AA$^{-1}$), with the parameters from Table \ref{table:params5}. 
Computing the absolute value of the dimensionless potential on the molecular surface gives over $55\%$ of the triangles with $\phi q_e/k_BT>1$ and an average value of 1.5.
Yet, using the linear or non-linear Poisson-Boltzmann model of APBS,\cite{BakerETal2001} the solvation energy for the isolated \ig 2a gives the same result.
(The APBS tests used a volumetric mesh of $150\times 150\times 150$\AA$^3$ with 449$^3$ nodes.)
Adding a surface with charge $\sigma=0.1$C/m$^2$ next to \ig 2a in the configuration of Fig.~\ref{fig:1IGT_3D_sig02_kap003125_til076-rot160}, the situation is similar: $51\%$ of the triangles have a dimensionless potential with absolute value exceeding 1 and the average is 1.3. By comparison with the isolated \ig 2 case, we expect linearized Poisson Boltzmann to give a good approximation of the solvation energy in this case.
This conclusion is also consistent with the results of Ref.~\onlinecite{Fogolari99} that show good agreement between linear and non-linear Poisson Boltzmann when the average dimensionless potential on the molecular surface is between $-2$ and $2$. 
Finally, with a value of surface charge of 0.1C/m$^2$, the mechanism of protein orientation appears to be dictated by the solvation energy, rather than the surface energy. That is, the maximum probability of the preferred orientation occurs at the minimum of solvation energy.\footnote{See Supplementary Materials.} Since solvation energy in our systems is well approximated by linear PB, we conclude that the use of this theory is justified in these cases.

\begin{figure*}
   \centering
   \subfloat[Probability distribution]{\includegraphics[width=0.4\textwidth]{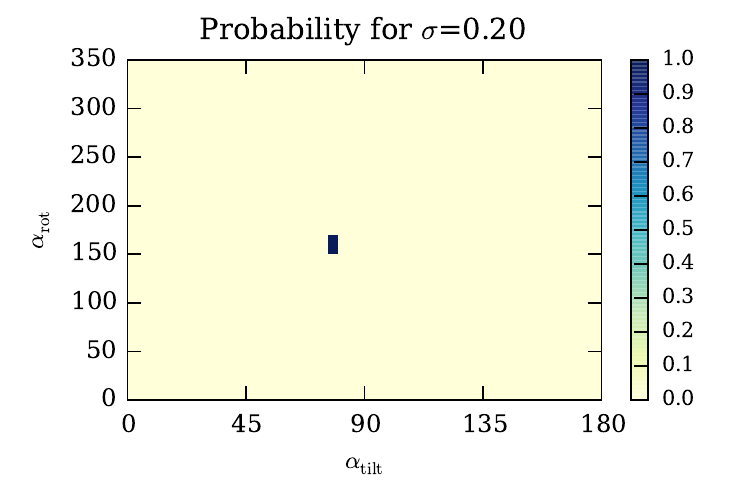} \label{fig:1IGT_sig02}}
   \subfloat[x-y plane view for $\alpha_{\text{tilt}} = 76^{\circ}$ and $\alpha_{\text{rot}} = 160^{\circ}$]{\includegraphics[width=0.4\textwidth]{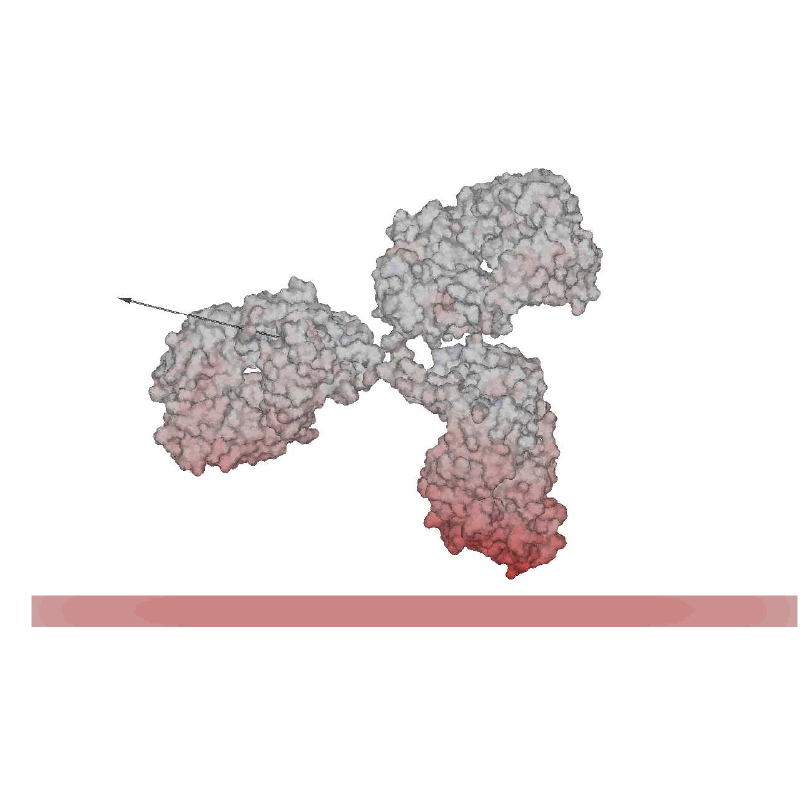} \label{fig:1IGT_3D_sig02}}
   \caption{Orientation probability distribution and surface potential of the preferred orientation for immunoglobulin G near a surface with $\sigma=0.2$C/m$^2$ and $\kappa=0.03125$\AA$^{-1}$. Note that these conditions are outside the range of linearized theory (as explained in the Discussion). }
   \label{fig:1IGT_sigma02}
\end{figure*}

Our results suggest that a combination of high surface charge and low salt concentration increases the probability of the preferred configuration, and that more favorable orientations are obtained with positive surface charge.
We completed an additional set of tests for the orientation of \ig 2a near a surface with $\sigma =0.2$C/m${^2}$ and $\kappa=0.03125$\AA$^{-1}$, even though the electrostatic potential is outside the linear regime in this case.
The result in Figure \ref{fig:1IGT_sigma02} suggests that as we increase the electrostatic effects, the preferred orientation becomes highly marked, and tends towards a favorable orientation for biosensors.
However, our model is based on the linearized Poisson-Boltzmann equation and this test goes outside the known regime where the model can be applied. We cannot claim that this is a physical situation, but it indicates a trend. Further study of the conditions that make \ig 2a orient favorably for biosensors may need nonlinear models, and combined experimental observations.

\section{Conclusion}
%!TEX root = CooperBarba-orientation.tex

Various studies have revealed the importance of protein orientation in immunoassays. One work suggested that highly oriented antibodies could result in 100$\times$ improvement in the affinity of a biosensor.\cite{TajimaTakaiIshihara2011} Thus, a design goal would be to know how to prepare a surface to control protein orientation. Yet, despite much work, control of protein orientation has not been successful.
This study increases our understanding of how nanosurface properties (charge) and preparation conditions (salt levels) affect protein orientation.
We successfully used an implicit-solvent model to study protein orientation near charged surfaces, which in our method can have any geometry. In a companion publication,\cite{CooperBarba2015a} we describe expanding the applicability of our open-source code, \pygbe, to account for the presence of charged surfaces and present grid-convergence studies using an analytical solution and protein \gb. 

Protein \gb behaves like a point dipole near a charged surface, with the dipole-moment vector shifting $\sim$180$^\circ$ when the sign of the surface charge flips. Our results compare well with experimental observations and simulations using combined Monte Carlo and molecular dynamics methods, supporting the use of our approach for probing protein orientation near charged surfaces.
We applied our approach to immunoglobulin G, a biomolecule that is much larger than protein \gb (about $125\times$, by volume) and would be challenging  to study via molecular dynamics. 
The iso-type \ig 2a was found by previous studies to be hard to control, exhibiting many orientations, but we are able to obtain a preferred orientation that is favorable for biosensing with a positive surface of 0.05C/m$^{2}$ or higher d 37mM of salt in the solvent. We conclude that local electrostatic interactions dominate over the dipole moment, and even this protein can be favorably oriented with the appropriate fabrication protocol. Potentially, protein engineering could be used to obtain ligand molecules that interact with charged surfaces in a desired fashion.
In this application, where ligand molecules undergo little conformational change as they adsorb on the sensor surface, our new implicit-solvent model can offer a valuable approach to assist in biosensor design. In our future work, and in collaboration with experimental researchers, we intend to use this approach to aid the design of better ligand molecules, by looking at the preferred orientations for different ligand mutants.

\begin{acknowledgments}
 This work was supported by ONR via grant \#N00014-11-1-0356 of the Applied Computational Analysis Program. LAB also acknowledges support from NSF CAREER award OCI-1149784 and from NVIDIA, Inc.\ via the CUDA Fellows Program. 
 We are grateful for many helpful conversations with members of the Materials and Sensors Branch of the Naval Research Laboratory, especially Dr. Jeff M. Byers and Dr. Marc Raphael.
\end{acknowledgments}

\bibliography{compbio,bem,scicomp,fastmethods,scbib,biosensors}

\end{document}